\RequirePackage{amsmath,lineno}
\documentclass[aps,prd,preprint,tightenlines,superscriptaddress,showpacs,byrevtex]{revtex4}  
\usepackage[]{hyperref}
\usepackage{lscape}
\usepackage{graphicx} 
\usepackage{dcolumn}  
\usepackage{bm}
\usepackage{amsmath}
\usepackage{color}
\usepackage{rotating}
\usepackage{cleveref}
\usepackage{lineno}

\crefname{equation}{Eq.}{Eqs.}
\crefname{figure}{Fig.}{Figs.}

\definecolor{darkgreen}{rgb}{0.,0.55,0.}
\definecolor{orange}{rgb}{0.75,0.6,0.}
\graphicspath{{ps}}

\newcommand{\Brprm}{\ensuremath{B^{0} \to \rho^{+}\rho^{-}}}
\newcommand{\Brr}{\ensuremath{B^{0} \to \rho^{0}\rho^{0}}}

\newcommand{\aonepi}{\ensuremath{B^{0} \to a^{\pm}_{1}\pi^{\mp}}}

\newcommand{\drho}{\ensuremath{B^{+} \to \bar{D}^{0} [K^{+}\pi^{-}\pi^{0}] \rho^{+}}}

\newcommand{\epem}{\ensuremath{e^{+} e^{-}}}
\newcommand{\qqbar}{\ensuremath{q \bar{q}}}
\newcommand{\BBbar}{\ensuremath{B \bar{B}}}
\newcommand{\BzBzb}{\ensuremath{B^{0} \bar{B}^{0}}}
\newcommand{\BpBm}{\ensuremath{B^{+} B^{-}}}

\newcommand{\pip}{\ensuremath{\pi^{+}}}
\newcommand{\pim}{\ensuremath{\pi^{-}}}
\newcommand{\piz}{\ensuremath{\pi^{0}}}

\newcommand{\rz}{\ensuremath{\rho^{0}}}
\newcommand{\rpm}{\ensuremath{\rho^{\pm}}}
\newcommand{\rop}{\ensuremath{\rho^{+}}}
\newcommand{\rom}{\ensuremath{\rho^{-}}}

\newcommand{\aone}{\ensuremath{a_{1}^{\pm}}}

\newcommand{\Ks}{\ensuremath{K^{0}_{S}}}
\newcommand{\Bz}{\ensuremath{B^{0}}}

\newcommand{\Ups}{\ensuremath{\Upsilon(4S)}}

\newcommand{\Btag}{\ensuremath{B^{0}_{\rm tag}}}

\newcommand{\Mbc}{\ensuremath{M_{\rm bc}}}
\newcommand{\De}{\ensuremath{\Delta E}}
\newcommand{\Fsb}{\ensuremath{{\cal F}_{S/B}}}
\newcommand{\fevt}{\ensuremath{{\cal F}_{S/B}}}
\newcommand{\mppz}{\ensuremath{m_{\pi^{\pm}\pi^{0}}}}

\newcommand{\mpz}{\ensuremath{m_{\pi^{+}\pi^{0}}}}
\newcommand{\mmz}{\ensuremath{m_{\pi^{-}\pi^{0}}}}

\newcommand{\Dt}{\ensuremath{\Delta t}}

\newcommand{\Mpp}{\ensuremath{m_{\pi^{+}\pi^{-}}}}
\newcommand{\cH}{\ensuremath{\cos\theta_{\rm H}}}

\newcommand{\phitwo}{\ensuremath{\phi_{2}}}
\newcommand{\phitwoeff}{\ensuremath{\phi^{\rm eff}_{2}}}

\begin{document}

\title{ \quad\\[0.5cm]  Study of $\bm{\Brprm}$ decays and implications
  for the CKM angle $\bm{\phi_2}$}

\noaffiliation
\affiliation{University of the Basque Country UPV/EHU, 48080 Bilbao}
\affiliation{Beihang University, Beijing 100191}
\affiliation{University of Bonn, 53115 Bonn}
\affiliation{Budker Institute of Nuclear Physics SB RAS, Novosibirsk 630090}
\affiliation{Faculty of Mathematics and Physics, Charles University, 121 16 Prague}
\affiliation{Chonnam National University, Kwangju 660-701}
\affiliation{University of Cincinnati, Cincinnati, Ohio 45221}
\affiliation{Deutsches Elektronen--Synchrotron, 22607 Hamburg}
\affiliation{Justus-Liebig-Universit\"at Gie\ss{}en, 35392 Gie\ss{}en}
\affiliation{Gifu University, Gifu 501-1193}
\affiliation{II. Physikalisches Institut, Georg-August-Universit\"at G\"ottingen, 37073 G\"ottingen}
\affiliation{SOKENDAI (The Graduate University for Advanced Studies), Hayama 240-0193}
\affiliation{Gyeongsang National University, Chinju 660-701}
\affiliation{Hanyang University, Seoul 133-791}
\affiliation{University of Hawaii, Honolulu, Hawaii 96822}
\affiliation{High Energy Accelerator Research Organization (KEK), Tsukuba 305-0801}
\affiliation{IKERBASQUE, Basque Foundation for Science, 48013 Bilbao}
\affiliation{Indian Institute of Technology Bhubaneswar, Satya Nagar 751007}
\affiliation{Indian Institute of Technology Guwahati, Assam 781039}
\affiliation{Indian Institute of Technology Madras, Chennai 600036}
\affiliation{Indiana University, Bloomington, Indiana 47408}
\affiliation{Institute of High Energy Physics, Chinese Academy of Sciences, Beijing 100049}
\affiliation{Institute of High Energy Physics, Vienna 1050}
\affiliation{Institute for High Energy Physics, Protvino 142281}
\affiliation{INFN - Sezione di Torino, 10125 Torino}
\affiliation{J. Stefan Institute, 1000 Ljubljana}
\affiliation{Kanagawa University, Yokohama 221-8686}
\affiliation{Institut f\"ur Experimentelle Kernphysik, Karlsruher Institut f\"ur Technologie, 76131 Karlsruhe}
\affiliation{Kennesaw State University, Kennesaw GA 30144}
\affiliation{King Abdulaziz City for Science and Technology, Riyadh 11442}
\affiliation{Department of Physics, Faculty of Science, King Abdulaziz University, Jeddah 21589}
\affiliation{Korea Institute of Science and Technology Information, Daejeon 305-806}
\affiliation{Korea University, Seoul 136-713}
\affiliation{Kyungpook National University, Daegu 702-701}
\affiliation{\'Ecole Polytechnique F\'ed\'erale de Lausanne (EPFL), Lausanne 1015}
\affiliation{Faculty of Mathematics and Physics, University of Ljubljana, 1000 Ljubljana}
\affiliation{Ludwig Maximilians University, 80539 Munich}
\affiliation{University of Maribor, 2000 Maribor}
\affiliation{Max-Planck-Institut f\"ur Physik, 80805 M\"unchen}
\affiliation{School of Physics, University of Melbourne, Victoria 3010}
\affiliation{Middle East Technical University, 06531 Ankara}
\affiliation{Moscow Physical Engineering Institute, Moscow 115409}
\affiliation{Moscow Institute of Physics and Technology, Moscow Region 141700}
\affiliation{Graduate School of Science, Nagoya University, Nagoya 464-8602}
\affiliation{Kobayashi-Maskawa Institute, Nagoya University, Nagoya 464-8602}
\affiliation{Nara Women's University, Nara 630-8506}
\affiliation{National Central University, Chung-li 32054}
\affiliation{National United University, Miao Li 36003}
\affiliation{Department of Physics, National Taiwan University, Taipei 10617}
\affiliation{H. Niewodniczanski Institute of Nuclear Physics, Krakow 31-342}
\affiliation{Niigata University, Niigata 950-2181}
\affiliation{Novosibirsk State University, Novosibirsk 630090}
\affiliation{Osaka City University, Osaka 558-8585}
\affiliation{Pacific Northwest National Laboratory, Richland, Washington 99352}
\affiliation{Peking University, Beijing 100871}
\affiliation{University of Pittsburgh, Pittsburgh, Pennsylvania 15260}
\affiliation{University of Science and Technology of China, Hefei 230026}
\affiliation{Soongsil University, Seoul 156-743}
\affiliation{Sungkyunkwan University, Suwon 440-746}
\affiliation{School of Physics, University of Sydney, NSW 2006}
\affiliation{Department of Physics, Faculty of Science, University of Tabuk, Tabuk 71451}
\affiliation{Tata Institute of Fundamental Research, Mumbai 400005}
\affiliation{Excellence Cluster Universe, Technische Universit\"at M\"unchen, 85748 Garching}
\affiliation{Department of Physics, Technische Universit\"at M\"unchen, 85748 Garching}
\affiliation{Toho University, Funabashi 274-8510}
\affiliation{Tohoku University, Sendai 980-8578}
\affiliation{Earthquake Research Institute, University of Tokyo, Tokyo 113-0032}
\affiliation{Department of Physics, University of Tokyo, Tokyo 113-0033}
\affiliation{Tokyo Institute of Technology, Tokyo 152-8550}
\affiliation{Tokyo Metropolitan University, Tokyo 192-0397}
\affiliation{University of Torino, 10124 Torino}
\affiliation{Utkal University, Bhubaneswar 751004}
\affiliation{CNP, Virginia Polytechnic Institute and State University, Blacksburg, Virginia 24061}
\affiliation{Wayne State University, Detroit, Michigan 48202}
\affiliation{Yamagata University, Yamagata 990-8560}
\affiliation{Yonsei University, Seoul 120-749}
 \author{P.~Vanhoefer}\affiliation{Max-Planck-Institut f\"ur Physik, 80805 M\"unchen} 
  \author{J.~Dalseno}\affiliation{Max-Planck-Institut f\"ur Physik, 80805 M\"unchen}\affiliation{Excellence Cluster Universe, Technische Universit\"at M\"unchen, 85748 Garching} 
 \author{C.~Kiesling}\affiliation{Max-Planck-Institut f\"ur Physik, 80805 M\"unchen} 
  \author{A.~Abdesselam}\affiliation{Department of Physics, Faculty of Science, University of Tabuk, Tabuk 71451} 
  \author{I.~Adachi}\affiliation{High Energy Accelerator Research Organization (KEK), Tsukuba 305-0801}\affiliation{SOKENDAI (The Graduate University for Advanced Studies), Hayama 240-0193} 
  \author{H.~Aihara}\affiliation{Department of Physics, University of Tokyo, Tokyo 113-0033} 
  \author{S.~Al~Said}\affiliation{Department of Physics, Faculty of Science, University of Tabuk, Tabuk 71451}\affiliation{Department of Physics, Faculty of Science, King Abdulaziz University, Jeddah 21589} 
  \author{K.~Arinstein}\affiliation{Budker Institute of Nuclear Physics SB RAS, Novosibirsk 630090}\affiliation{Novosibirsk State University, Novosibirsk 630090} 
  \author{D.~M.~Asner}\affiliation{Pacific Northwest National Laboratory, Richland, Washington 99352} 
  \author{H.~Atmacan}\affiliation{Middle East Technical University, 06531 Ankara} 
% ...
\collaboration{The Belle Collaboration}

\begin{abstract}
We present a measurement of the branching fraction and the longitudinal polarization fraction of $\Brprm$ decays, as well as the time-dependent $CP$ violating parameters in decays into longitudinally polarized $\rop\rom$ pairs with Belle's final data set of $772\times 10^6$ $B\bar{B}$ pairs, at the \Ups\ resonance, collected at the asymmetric-energy $e^+ e^-$ collider KEKB. We obtain
  \begin{eqnarray}                                                                                                                                      
{\cal B}(\Brprm)&=&(28.3\pm 1.5\;(\rm stat) \pm 1.5\;(\rm syst))\times 10^{-6}, \nonumber\\ 
f_L &=& 0.988\pm 0.012\;(\rm stat ) \pm0.023\;(\rm syst),\nonumber\\          
  {\cal A}_{CP} &=&0.00\pm0.10\;(\rm stat ) \pm0.06\;(\rm syst), \nonumber\\
 {\cal S}_{CP} &=&-0.13\pm0.15\;(\rm stat ) \pm0.05\;(\rm syst).\nonumber        \end{eqnarray}                                                                                                                                       
We perform an isospin analysis to constrain the CKM angle $\phi_2$ and obtain two solutions with      
 \begin{eqnarray}    
 \phi_{2} = (93.7\pm10.6)^{\circ},\nonumber   
 \end{eqnarray}                                                                                                                      
 being most compatible with other Standard-Model based fits to the data. 

\end{abstract}


\maketitle

\tighten


\section{Introduction}
\begin{figure}[b]
  \centering
  \includegraphics[height=120pt,width=!]{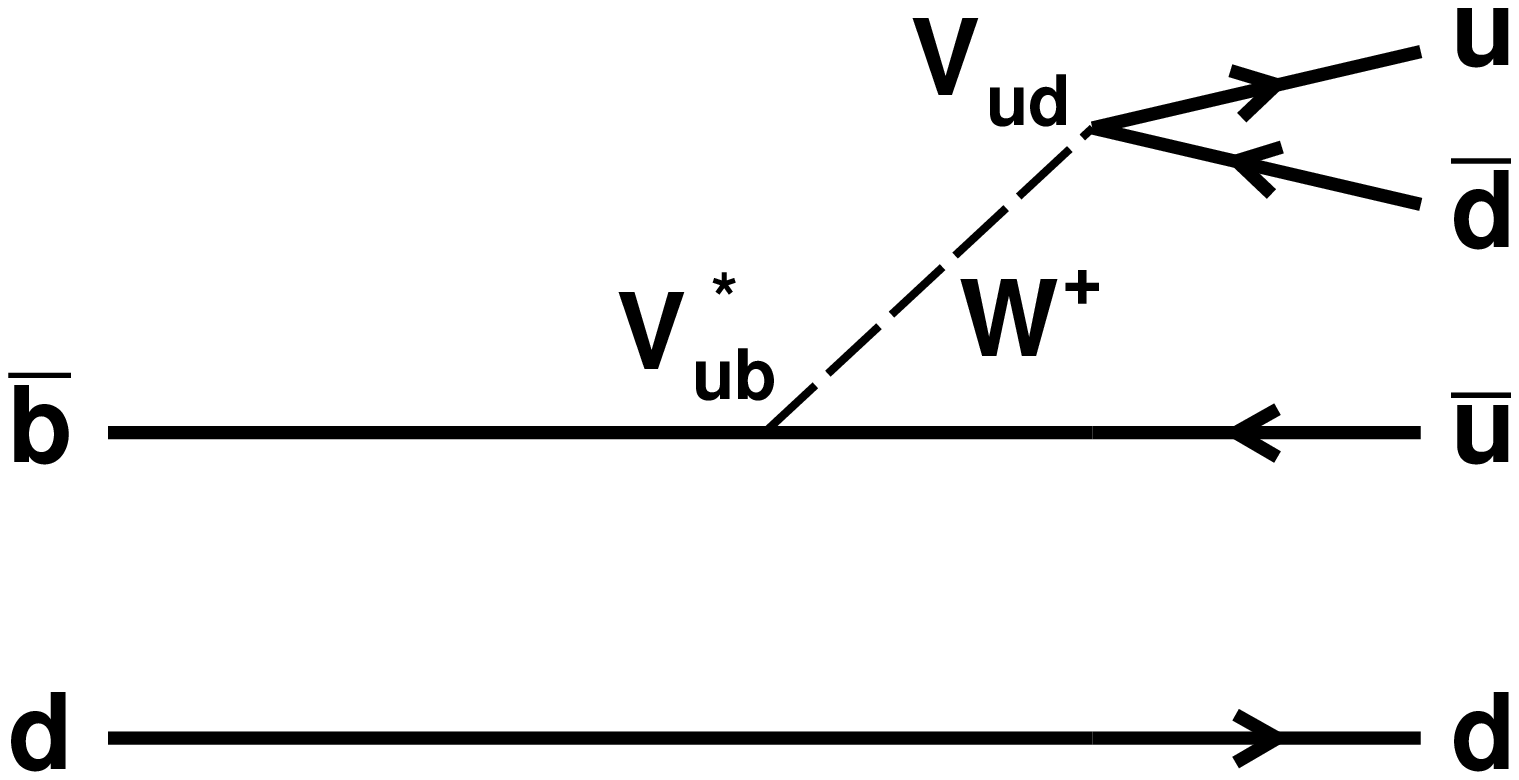}
  \includegraphics[height=120pt,width=!]{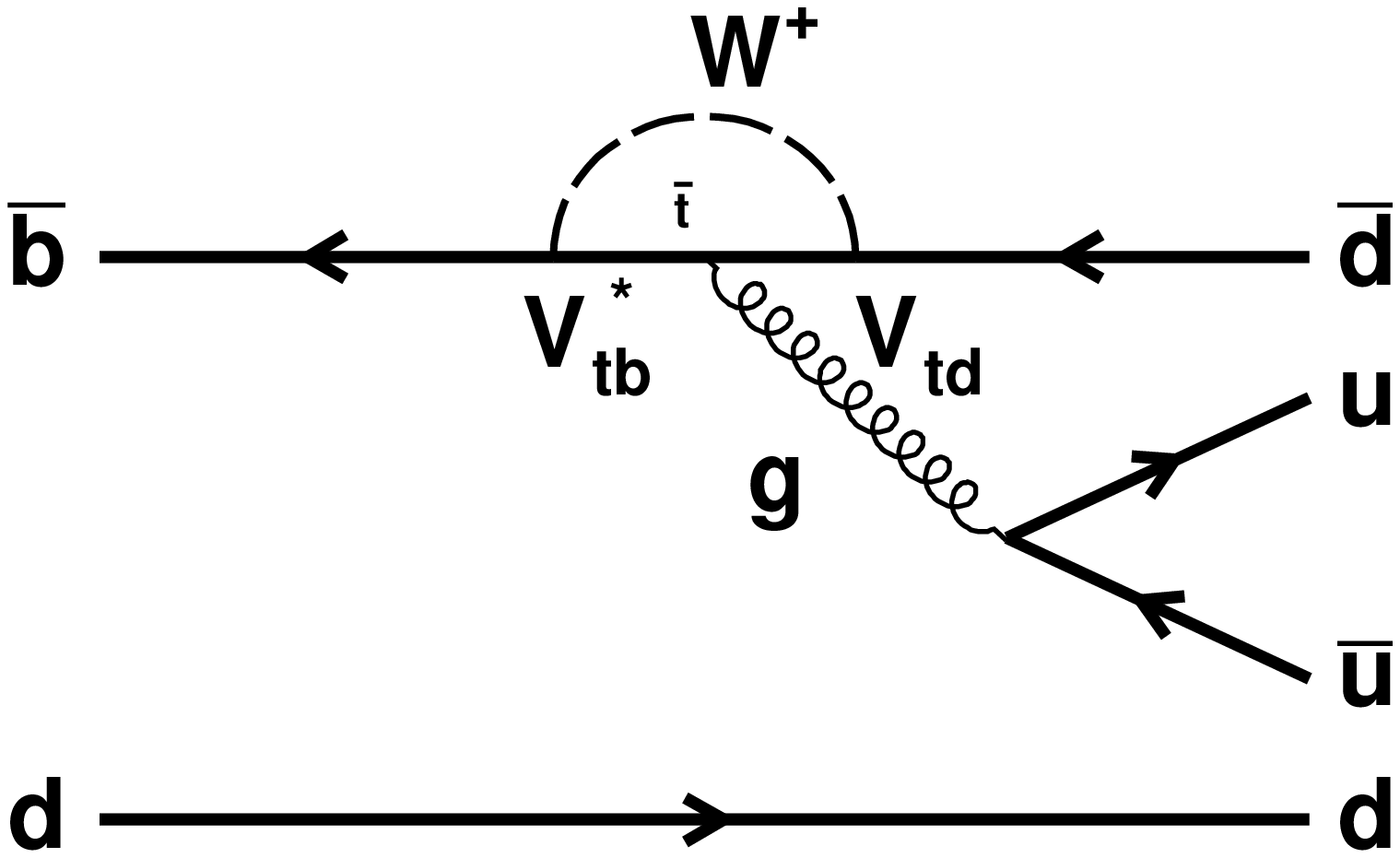}
  \caption{Leading-order tree (left) and penguin (right) diagrams for the decay \Brprm.}
  \label{fig_r0r0}
\end{figure}

$CP$ violation in the Standard Model (SM) is due to an irreducible complex phase in the Cabibbo-Kobayashi-Maskawa (CKM) quark-mixing matrix~\cite{Cabibbo,KM}. Mixing-induced $CP$ violation in the $B$ sector has been clearly observed by the Belle~\cite{jpsiks_Belle, jpsiks_Belle2} and BaBar~\cite{jpsiks_BABAR, jpsiks_BABAR2} collaborations in the $b \rightarrow c \bar c s$ transition~\cite{CCimplied} in $\Bz \rightarrow J/\psi \Ks$, while many other modes provide additional information on $CP$ violating parameters~\cite{BelleResults, CKMandUT}.\\
At the \Ups\ resonance, a quantum-entangled $\BzBzb$ pair is produced via $\epem\to\Ups\to\BzBzb$. When one of the two $B$ mesons ($B^{0}_{CP}$) decays into the $CP$ eigenstate of interest at time $t_{CP}$, the flavor $q$ of the other $B$ meson ($B^{0}_{\rm tag}$, decaying at time $t_{\rm tag}$) determines the flavor of $B^{0}_{CP}$ at the latter time: $q=+1$ for $B^{0}_{\rm tag}=B^0$ and $q=-1$ for $B^{0}_{\rm tag}=\bar{B}^0$. The time interval between the decays of the two $B$ mesons is defined as $\Delta t\equiv t_{CP}-t_{\rm tag}$ and the time-dependent rate for a $B$ decay into a $CP$ eigenstate is given by
\begin{equation}
{\cal P}(\Delta t, q)=\frac{e^{-|\Delta t|/\tau_{B^{0}}}}{4\tau_{B^{0}}}\bigg[1+q\bigg({\cal A}_{CP}\cos(\Delta m \Delta t) + {\cal S}_{CP}\sin(\Delta m \Delta t)\bigg)\bigg].
\label{e_dtq}
\end{equation}
Here, $\tau_{\Bz}$ is the $B^0$ lifetime and $\Delta m$ the mass difference of the two mass eigenstates of the neutral $B$ meson. ${\cal A}_{CP}$ and ${\cal S}_{CP}$ are the observables for direct and mixing-induced $CP$ violation, respectively. \\
In this measurement, we extract the branching fraction ${\cal B}$, the fraction of longitudinal polarization of the $\rho$ mesons and the $CP$-violating parameters in $B^0(\bar{B}^0)\to \rho^+\rho^-$ decays, also referred to as ``signal''. The $CP$-violating parameters ${\cal A}_{CP}$ and ${\cal S}_{CP}$ are measured only for decays into longitudinally polarized $\rho$ mesons. The leading-order tree and penguin diagrams of $\Brprm$ decays are shown in~\Cref{fig_r0r0}. These decays proceed predominantly through the $\bar{b} \rightarrow \bar{u} u \bar{d}$ transition and are therefore sensitive to one of the internal angles of the roughly equilateral unitarity triangle, $\phitwo\;({\rm or}\;\alpha) \equiv \arg[(-V_{td}V^{*}_{tb})/(V_{ud}V^{*}_{ub})]$; its current world average is $(87.7^{+3.5}_{-3.3})^{\circ}$~\cite{ckmfitter}. The Belle, BaBar and LHCb collaborations have reported time-dependent $CP$ asymmetries in the following modes: $\Bz \rightarrow \pip \pim$~\cite{pipi_Belle,pipi_BABAR,pipi_LHCb}, $\rho^{\pm} \pi^{\mp}$~\cite{rhopi_Belle,rhopi_BABAR}, $\rho^{+} \rho^{-}$~\cite{Brprm_Belle,rhorho_Belle,rhorho_BABAR}, $\rho^{0} \rho^{0}$~\cite{Br0r0_pit,r0r0_BABAR} and $a_{1}^{\pm}\pi^{\mp}$~\cite{jeremy_a1pi,a1pi_BABAR1,a1pi_BABAR2}. A feature common to these measurements is that possible loop contributions, in addition to the leading-order tree amplitude, can shift the measured angle to $\phitwoeff\ \equiv  \phitwo\ + \Delta\phi_{2}$, so that the observed mixing-induced $CP$-violation parameters are related by ${\cal S}_{CP}=\sqrt{1-{\cal A}_{CP}^{2}}\sin(2\phitwoeff)$. This inconvenience can be overcome by estimating $\Delta \phitwo$ using either an isospin analysis~\cite{theory_isospin} or $SU(3)$ flavor symmetry~\cite{theory_SU3}.

\begin{figure}[b]
  \centering
  \includegraphics[height=140pt,width=!]{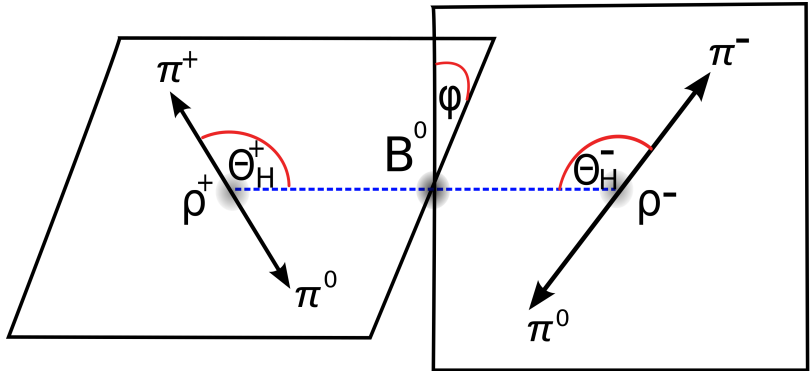}
  \caption{Definition of the helicity angles $\theta_{\rm H}^\pm$ for each $\rho^\pm$, identified by its charge.}
  \label{fig_helicity}
\end{figure}

The $\rop\rom$ vector-vector state is a superposition of three helicity amplitudes $A_{0}, A_{-1}$ and $A_{+1}$ with $CP$-even and -odd contributions. Their fractions can be determined through an angular analysis; the $\rho^\pm$ mesons from \Brprm\ are found to be almost entirely longitudinally polarized~\cite{Brprm_Belle,rhorho_BABAR}. We use the helicity basis, which allows us to separate longitudinally ($CP$-even, with amplitude $A_{0}$) from transversely ($CP$-even and -odd, with amplitudes $A_{\pm}$) polarized $\rho$ mesons. The distribution of the two angles $\theta_{\rm H}^{+}$ and $\theta_{\rm H}^{-}$, each defined as the angle between one of the daughters of the $\rho^\pm$ meson (here, the charged pion) and the $B$ flight direction in the corresponding rest frame of the $\rho^\pm$ (see ~\Cref{fig_helicity}), is sensitive to the polarization:
\begin{equation}
\frac{1}{\Gamma}\frac{d^{2}\Gamma}{d\cos\theta_{\rm H}^{+}d\cos\theta_{\rm H}^{-}} = \frac{9}{4}\biggl[\frac{1}{4}(1 - f_L)\sin^{2}\theta_{\rm H}^{+}\sin^{2}\theta_{\rm H}^{-} + f_L\cos^2\theta_{\rm H}^{+}\cos^2\theta_{\rm H}^{-}\biggr],
\label{e_helicity}
\end{equation}
 where $f_L = |A_0|^{2}/\sum |A_{i}|^{2}$ is the fraction of longitudinal polarization. \\
The SM, using perturbative QCD (pQCD) or QCD factorization in the heavy-quark limit~\cite{pQCD, fQCD2, fQCD3, fQCD4, BVV, fQCD, BVV2,Bm1m2_NNLO}, predicts the \Brprm\ branching fraction to be $\sim 30\times 10^{-6}$ and $f_L\sim 1$. Furthermore, no direct $CP$ violation is expected if penguin contributions are found to be small. The previous measurements are summarized in~\Cref{tab_prev_meas}. The main improvements here compared to previous Belle measurements are the increased data sample and the simultaneous extraction of all observables. The inclusion of additional observables in the fit improves the signal-to-background discrimination and allows us to relax selection criteria and consequently increase the signal efficiency.

\begin{table}[h]
\centering
\caption{Previous measurements of $B\to\rho^+\rho^-$ decays. The first error is
  statistical and the second is systematic. The Belle results on the branching fraction and $f_L$ (the $CP$-violating parameters) are obtained from $275\times10^6$ ($384\times 10^6$) \BBbar\ pairs~\cite{Brprm_Belle,rhorho_Belle}. The BaBar results are obtained from $384\times 10^6$ \BBbar\ pairs~\cite{rhorho_BABAR}.}
   \begin{tabular}
      {@{\hspace{0.25cm}}c@{\hspace{0.25cm}} @{\hspace{0.25cm}}c@{\hspace{0.25cm}}  @{\hspace{0.25cm}}c@{\hspace{0.25cm}} @{\hspace{0.25cm}}c@{\hspace{0.25cm}}  @{\hspace{0.25cm}}c@{\hspace{0.2cm}}}      \hline \hline
    Exp.  & ${\cal B}[\times 10^{-6}]$ & $f_L$ & ${\cal A}_{CP}$ & ${\cal S}_{CP}$ \\[1ex]
\hline
Belle &$22.8\pm 3.8 \pm2.6$ &$0.94\pm 0.04 \pm0.03$  & $0.16\pm 0.21 \pm0.07$& $0.19\pm 0.30 \pm0.07$  \\   [1ex] 

\hline
BaBar &$25.5\pm2.1^{+3.6}_{-3.9}$ & $0.992\pm0.024^{+0.026}_{-0.013}$& $-0.01\pm0.15\pm0.06$ & $-0.17\pm0.20^{+0.05}_{-0.06}$     \\[1ex] 
   \hline \hline
    \end{tabular}
\label{tab_prev_meas}
 \end{table}

In~\Cref{Data Set And Belle Detector}, we describe briefly the data set and the Belle detector. The event selection and the model used for the measurement are described in~\Cref{Event Selection,Event Model}, respectively. The results are presented in~\Cref{Fit Result}, followed by validity checks in~\Cref{Validity Checks and Significance}. The systematic uncertainties are discussed in~\Cref{Systematic Uncertainties}. Constraints on the CKM phase \phitwo\ are presented in~\Cref{phi2 constraint}.

\section{Data Set And Belle Detector}
\label{Data Set And Belle Detector}
This measurement is based on the final data sample containing $772 \times 10^{6}$ \BBbar\ pairs collected with the Belle detector at the KEKB asymmetric-energy \epem\ ($3.5$ on $8~{\rm GeV}$) collider~\cite{KEKB}. At the \Ups\ resonance ($\sqrt{s}=10.58$~GeV), the Lorentz boost of the produced \BBbar\ pairs is $\beta\gamma=0.425$ along the $z$ direction, which is opposite to the positron beam direction.
In addition, $100 \;{\rm fb}^{-1}$ of data about $60$ MeV below the \Ups\ resonance threshold have been recorded and are referred to as ``off-resonance'' data.

The Belle detector is a large-solid-angle magnetic
spectrometer that consists of a silicon vertex detector (SVD),
a 50-layer central drift chamber (CDC), an array of
aerogel threshold Cherenkov counters (ACC), 
a barrel-like arrangement of time-of-flight
scintillation counters (TOF), and an electromagnetic calorimeter 
(ECL) comprised of CsI(Tl) crystals located inside 
a superconducting solenoid coil providing a 1.5~T
magnetic field.  An iron flux-return yoke located outside of
the coil is instrumented to detect $K_L^0$ mesons and to identify
muons.  The detector
is described in detail elsewhere~\cite{Belle}.
Two inner detector configurations were used: a 2.0 cm radius beampipe
and a 3-layer silicon strip vertex detector (SVD1) were used for the first sample
of $152 \times 10^6$ \BBbar\ pairs and a 1.5 cm radius beampipe, a 4-layer
silicon strip vertex detector (SVD2)~\cite{svd2} and a small-cell inner drift chamber for the remaining $620 \times 10^6$ \BBbar\ pairs. We use a GEANT-based~\cite{GEANT}
 Monte Carlo (MC) simulation to model the response of the detector and to determine
 its acceptance.

\section{Event Selection}
\label{Event Selection}
We reconstruct $\Brprm$, where $\rho^\pm\to\pi^\pm\pi^0$ and $\piz \to \gamma\gamma$. Charged tracks must satisfy requirements on the distance of closest approach to the interaction point: $|dz| < 5.0 \; {\rm cm}$ and $dr < 0.5 \; {\rm cm}$ along and perpendicular to the $z$ axis, respectively. We select charged pions based on particle identification (PID) information from the CDC, ACC and TOF. Our PID requirement on the kaon-pion separation retains 90\% of all pions from \Brprm\ decays, but only 10\% of all kaons. In addition, we use information from the ECL to veto particles consistent with the electron hypothesis. Requirements of at least two SVD hits in the $z$ projections and one in the azimuthal~\cite{ResFunc} are imposed on the charged tracks.
A $\piz$ candidate is reconstructed from two photons, identified by isolated energy clusters in the ECL. We suppress combinatorial background by requiring a minimum photon energy of $E_\gamma>50 \;(90) {\;\rm MeV}$ in the ECL barrel (endcap) region and require that the invariant mass of the photon pair to be near the $\piz$ mass: $|m_{\gamma\gamma}-m_{\pi^0}|<15 {\;\rm MeV/{\rm c}^2}$, which covers about three times the experimental resolution. We perform a $\piz$ mass-constraint fit and retain $\pi^0$ candidates fulfilling $\chi^2<50$ and $p_{\piz}^*>100 \;{\rm MeV/{\rm c}}$, where $p_{\piz}^*$ is the momentum of the $\piz$ in the center-of-mass system (CMS).

Intermediate charged-dipion states are reconstructed within the invariant-mass range $0.4 \; {\rm GeV}/c^{2} < m(\pi^\pm \piz) < 1.15 \; {\rm GeV}/c^{2}$, covering the broad $\rho^\pm(770)$ resonance~\cite{PDG}. This retains $92\%$ of the phase space available for two $\rho^\pm$ mesons from $\Brprm$ decays, while reducing combinatorial background with a tendency for higher dipion masses. Upon combination of two dipion states with opposite charge, a $\Brprm$ candidate is formed. All remaining particles are associated with the accompanying \Btag\ meson.

Reconstructed $B$ candidates are described by two kinematic variables: the beam-energy-constrained mass $\Mbc \equiv \sqrt{(E^{\rm CMS}_{\rm beam}/c^2)^{2} - (p^{\rm CMS}_{B}/c)^{2}}$ and the energy difference $\De \equiv E^{\rm CMS}_{B} - E^{\rm CMS}_{\rm beam}$, where $E^{\rm CMS}_{\rm beam}$ is the beam energy and $E^{\rm CMS}_{B}$ ($p^{\rm CMS}_{B}$) is the energy (momentum) of the $B$ meson, evaluated in the CMS. $B$ candidates satisfying  $\Mbc > 5.27 \; {\rm GeV}/c^{2}$  and $|\De| < 0.15 \; {\rm GeV}$ are selected for further analysis.

\begin{figure}[h]
  \centering
  \includegraphics[height=300pt,width=!]{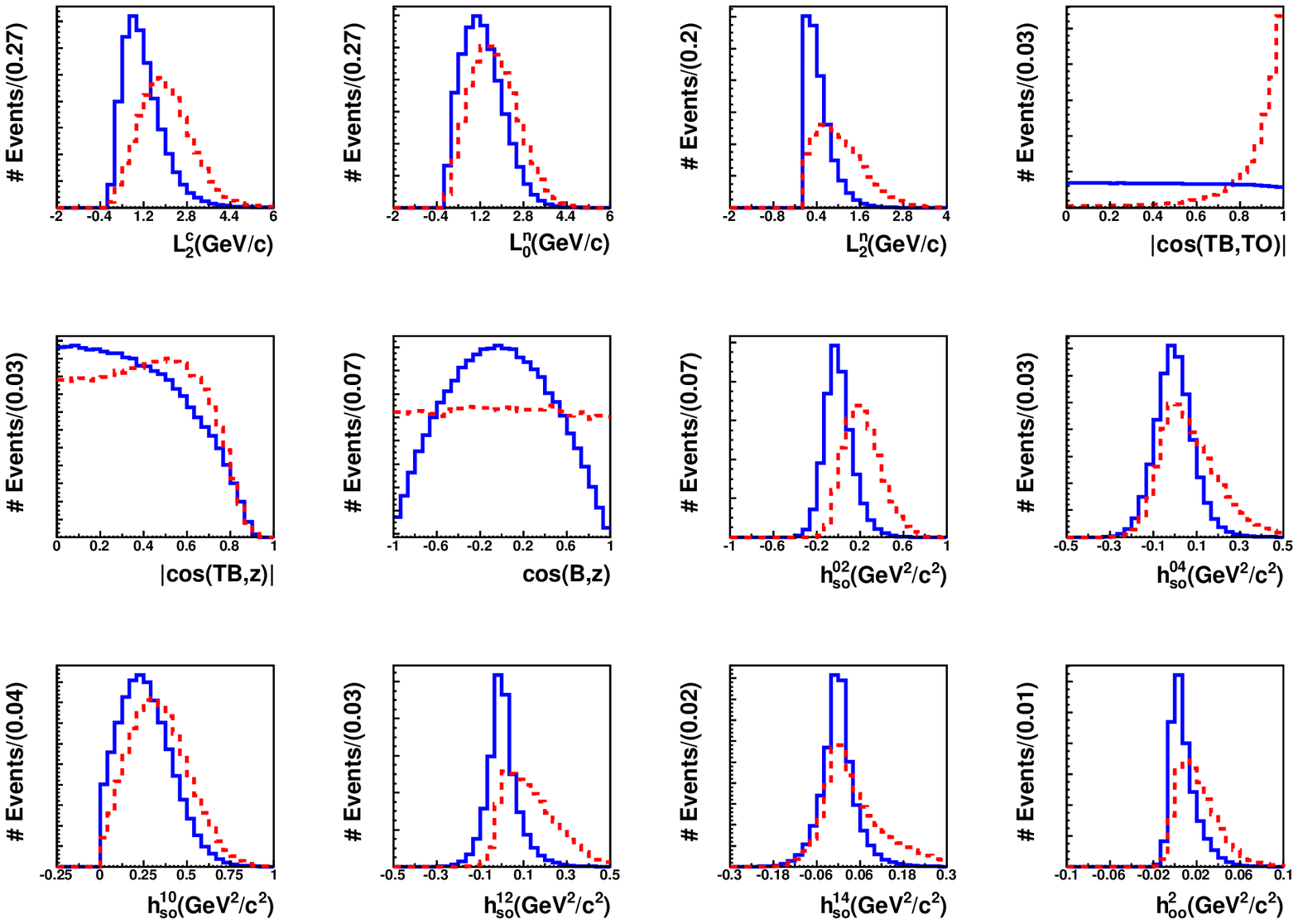} 
\put(-430, 270){(a)}\\
  \includegraphics[height=200pt,width=!]{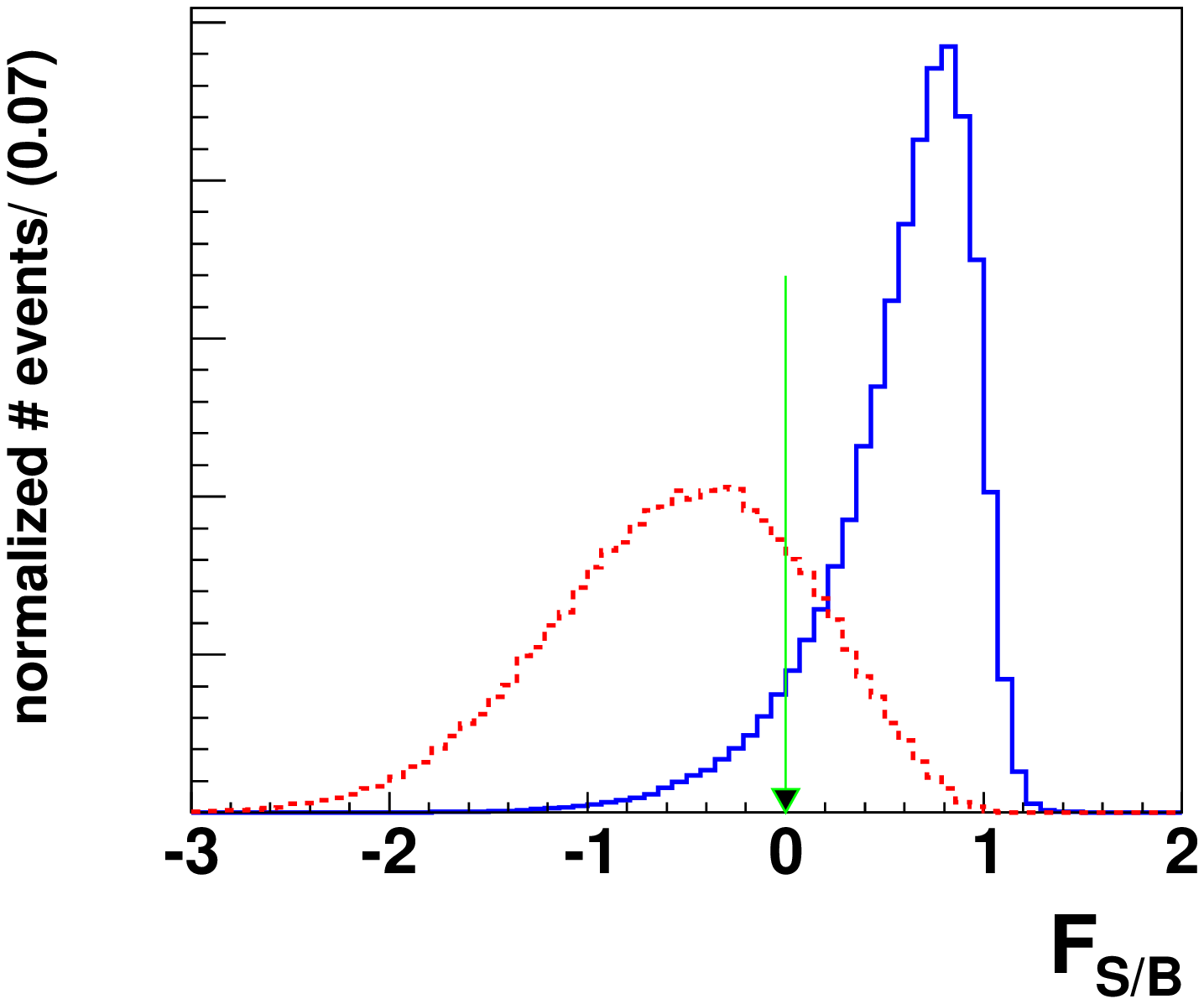}
\put(-320, 160){(b)}\\
  \caption{(color online) (a) Simulated MC and off-resonance data distributions for the quantities used
to construct the Fisher discriminant \Fsb. The solid (blue) histograms
show the distribution for \BBbar\ MC events, while the dashed (red)
histograms show the distributions for events from off-resonance data,
both normalized to the same area.\\
(b) Output of the fisher discriminant. The same line (color) scheme is used
as above. The green arrow indicates the requirement of $\fevt\ > 0$.}
  \label{fig_fd}
\end{figure}

The dominant background contribution arises from continuum events ($\epem \rightarrow \qqbar$, where $q=u,d,s,c$). We use their jet-like topology to separate them from the more spherical \BBbar\ decays using a Fisher discriminant~\cite{fisher} \Fsb, constructed from the following 12 variables (all evaluated in the CMS):
 \begin{itemize}
  \item $L_{0}^{\rm n}$, $L_{2}^{\rm n}$,  $L_{2}^{\rm c}$,\\
where $L_{i}^{\rm k} = \sum_j |\vec{p_{j}}|(\cos{\theta_{j}})^{i}$ for neutral clusters (k=n) and charged tracks (k=c) belonging to the tag side, where $i=0$ or $2$, $p_j$ is the momentum of the $j$-th particle and $\theta_{j}$ the angle between its direction and the thrust axis of the $B$ candidate~\cite{rhorho_BABAR}. 
 \item $|\cos(TB,TO)|$,\\
cosine of the angle between the thrust axis of the $B$ candidate ($TB$) and the thrust axis of the remaining tracks ($TO$).
 \item $|\cos(TB,z)|$,\\
cosine of the angle between $TB$ and the $z$-axis.
  \item $\cos(B,z)$,\\
the projection of the $B$ flight direction onto the $z$-axis,
\item[]and the following variables, closely related to the modified Fox-Wolfram moments~\cite{foxw1,foxw2}:
  \item $h_{so}^{c2}, h_{so}^{c4}, h_{so}^{n0}, h_{so}^{n2}, h_{so}^{n4}, h_{oo}^{2}$, \\
with $h_{so}^{km}=\sum\limits_{i,j}|\vec{p}_{j_k}| P_{m}(\cos\theta_{ij_{k}})$, where $\vec{p}_{j_{k}}$ (here and in the rest of this bullet) is the momentum of the $j_k$-th particle from the other side ($o$), the subscript $i$ labels the $i$-th track from the signal side ($s$), $\theta_{ij_{k}}$ is the angle between particles $i$ and $j_k$, and $P_m$ is Legendre polynomial of order $m$. For $o$, $k=c$ for charged tracks and $k=n$ for neutral particles, respectively. The quantity $h_{oo}^{2}=\sum\limits_{i,j}|\vec{p}_i||\vec{p}_{j}| P_{2}(\cos\theta_{ij})$ uses only particles from $o$ and does not consider their charge.
 \end{itemize}
 The respective distributions and the output are shown in~\Cref{fig_fd}. We require $\Fsb > 0$ to reject 80\% of the continuum background while retaining 80\% of signal. We use samples with signal MC events and off-resonance data taken below the \Ups\ resonance for the training of the Fisher discriminant. The requirement on $\Fsb$ together with the previously mentioned requirements of $\De$, $\Mbc$, and $m(\pi^\pm \piz)$ and the cuts $-0.85\leq \cH^\pm\leq 0.98$ and $|\Dt|<70 \;{\rm ps}$, define the fit region. The cut of the helicity angles reduces combinatorial background peaking at $\cH^\pm\to \pm 1$; the \Dt\ range is also used in independent studies to determine the systematic uncertainties related to the modeling of the $\Dt$ distributions.

According to signal MC simulation, $29\%$ of all events have multiple $\Brprm$ candidates. Selecting the $B$ candidate with the smallest sum of the $\chi^2$s from the $\piz$ mass constraint fits yields the correct $B$ in $79\%$ of all events with multiple candidates. If both possible dipion combinations of the four pions fall within the fit region, we choose the combination with the larger momentum difference between the daughter pions: longitudinally polarized $\rho$ mesons tend to decay into a high- and low-momentum pion in the CMS frame.

The vertices of the $B$ candidates are determined from their charged
tracks~\cite{ResFunc} and we employ the flavor-tagging method
described in Ref.~\cite{Tagging}. The tagging information is
represented by two parameters: the \Btag\ flavor $q=\pm1$, and the
tagging quality $r$, a continuous, event-by-event flavor tagging dilution factor
determined from MC simulation that ranges from zero for no flavor
discrimination to unity for unambiguous flavor assignment. We divide
the data into seven $r$-bins, labeled by the index $l$.  The
mistagging probability in each $r$-bin, $w$, is obtained from studying a high statistics
control sample. Thus, the $CP$ asymmetry in data is diluted by a factor 
$1-2w$ instead of the MC-determined $r$. The performance of the
flavor tagging algorithm is then given by $\epsilon_{\rm tag} =
(1-2w)^2\epsilon_{\rm raw}$, where $\epsilon_{\rm raw}$ is the raw
tagging efficiency. $\epsilon_{\rm tag}$ has been measured to $0.284\pm0.010$ for
SVD1 and $0.301\pm0.004$ for SVD2~\cite{jpsiks_Belle2}.
We find \Fsb\ to be correlated with $r$ and provide individual descriptions of \Fsb\ in each $r$-bin.

Since the distributions of the fit variables can significantly depend on the number of correctly assigned pions to the reconstructed $\rho^\pm$ meson candidates, we consider four categories of reconstruction quality: 
\begin{itemize}
\item[] a) ``truth'': all four pions correctly reconstructed, 
\item[] b) ``2T'': both charged pions correctly reconstructed and at least one \piz\ incorrectly assigned, 
\item[] c) ``1T'': only one correctly reconstructed charged pion and no requirement on the $\piz$ mesons, 
\item[] d) ``0T'': anything else.
\end{itemize}
The total reconstruction efficiencies (percentage of reconstructed events
with respect to all generated events) and their decompositions $\epsilon_i$ for both polarization states are obtained from MC simulation and are listed in~\Cref{tab_recEff}, where, for transverse polarization (TP), the $0T$ fraction is small and so is included in the $1T$ entry. All categories provide useful information for the measurement of the branching fraction and the polarization, while those with at least one correctly reconstructed $\pi^\pm$ are used for the time-dependent measurement.

\begin{table}[h]
\centering
\caption{Reconstruction efficiencies (all numbers in percent) for longitudinal (LP) and
  transverse (TP) polarizations obtained from fully simulated
  signal MC events.}
   \begin{tabular}
      {@{\hspace{0.5cm}}c@{\hspace{0.5cm}} @{\hspace{0.5cm}}c@{\hspace{0.5cm}} @{\hspace{0.5cm}}c@{\hspace{0.5cm}} @{\hspace{0.5cm}}c@{\hspace{0.5cm}} @{\hspace{0.5cm}}c@{\hspace{0.5cm}}| @{\hspace{0.5cm}}c@{\hspace{0.5cm}}   @{\hspace{0.5cm}}c@{\hspace{0.5cm}}}
      \hline \hline 
 Mode & truth & $2T$ & $1T$ & $0T$ & Total\\
   \hline \hline 
    $\epsilon_{\rm Sig}^{\rm SVD1\; (LP)}\;[\%]$  &  $5.2$ &$1.9$ & $1.3$ & $0.1$ & $8.5$\\[1ex] 
  $\epsilon_{\rm Sig}^{\rm SVD2\; (LP)}\;[\%]$ & $6.0$   & $2.2$ & $1.5$ & $0.2$ & $9.9$\\[1ex] 
\hline
   $\epsilon_{\rm Sig}^{\rm SVD1\; (TP)}\;[\%]$ & $12.0$ & $1.5$ & $0.4$ & $--$ &$13.9$\\[1ex] 
  $\epsilon_{\rm Sig}^{\rm SVD2\; (TP)}\;[\%]$ & $13.4$ & $1.7$ & $0.4$ & $--$ & $15.5$\\  [1ex] 
      \hline \hline
  \end{tabular}
\label{tab_recEff}
 \end{table}

\section{Event Model}
\label{Event Model}
After applying the selection criteria to the data, more than $99\%$ of
all events belong to background processes. The branching fraction,
$\rho$ polarization fraction and $CP$-violating parameters of \Brprm\
decays are extracted using an extended unbinned maximum likelihood fit
to \De, \Mbc, \Fsb, $M_+$, $M_-$, $H_+$, $H_-$, $\Dt$ and $q$ in the
$l^{\rm th}$ $r$-bin and SVD configuration $s$, where $M_\pm$ and
$H_\pm$ represent the invariant dipion mass $m(\pi^\pm\piz)$ and
helicity parameter $\cH^\pm$ of the $\rho^\pm$ candidate with
corresponding charge, respectively. Besides signal, we consider
continuum, four different combinatorial backgrounds from other $B$
decays (neutral and charged $B$ decays into charm and charmless final
states) and seven $B$ decay modes with a $\pip\piz\pim\piz$ final
state: $B^{0} \to  a_1^{\pm}\pi^{\mp}, \;a_1^{0}\pi^{0},\;\omega\piz,$
and the non-resonant final states $(\rho^\pm\pi^\mp\piz)_{\rm
  NR},\;(\rz\piz\piz)_{\rm NR},\; (f_0(980)\piz\piz)_{\rm NR},$ and
$(\pip\piz\pim\piz)_{\rm NR}$. These four-pion final states are
referred to as peaking backgrounds since their \De\ and \Mbc\
distributions mimic the signal. For the signal and background
components, probability density functions (PDFs) are defined with
parameters adjusted to describe the MC distributions for each of the
nine observables in the $l,s$ data subsamples. The PDF for event $i$
is given by ${\cal
  P}(\De^{i},\Mbc^{i},\Fsb^{i},M_+^{i},M_-^{i},H_+^{i},H_-^{i},\Dt^{i},q^{i})$.
Correlations between or among observables are taken into account by
incorporating a dependence of one variable's PDFs parameters on the
correlated variable(s). The component's PDF is taken as the
product of individual PDFs for each fit variable ${\cal P}(j)$. 
A summary of the model including all correlations that are accounted
for is given in~\Cref{t_model_suma} and
a detailed and complete description of the model is provided in Ref.~\cite{pit_phd}.\\
In analogy to the signal model, we consider up to three different reconstruction categories for background processes where $\rho^\pm$ resonances can be reconstructed: both ($2T$), one ($1T$), or no charged pion ($0T$) originating from the $\rho^\pm$ resonance.
In the case of only one correct track ($1T$), the distributions of the
dipion masses and helicity angles depend strongly on the charge of the
correctly reconstructed track (see e.g. Eq.\ref{model1T}). We account
for this in the description of the likelihood by considering each case
(\rop\ or \rom) separately, but symmetrizing the PDFs such that the
PDFs with a correct $\rop$ are identical to those with a correct
$\rom$ when interchanging the label ($+\leftrightarrow -$). All
relevant components are found to be charge symmetric and have equal
fractions of positively and negatively charged $\rho^\pm$ mesons. \\ 
The dipion resonances are described by a relativistic Breit-Wigner (BW)
\begin{equation}
  BW(m_{\pi\pi})  \equiv \frac{m_{0}\Gamma(m_{\pi\pi})}{(m_{\pi\pi}^{2}-m_{0}^{2})^{2} + m_{0}^{2}\Gamma^{2}(m_{\pi\pi})},
\label{e_bw}
\end{equation}
with a mass-dependent width
\begin{equation}
\Gamma(m_{\pi\pi}) = \Gamma_{0} \biggl(\frac{p_{\pi}}{p_{0}}\biggr)^{3} \biggl(\frac{m_{0}}{m_{\pi\pi}}\biggr)B^{2}_{1}(p_{\pi}),
\end{equation}
where $p_{\pi}$ is the momentum of either resonance daughter in the resonance frame and $m_{\pi\pi}$ is the invariant mass of the dipion pair. $\Gamma_{0}$ and $m_{0}$ are the width and mass of the nominal resonance and $p_{0}$ is the nominal momentum of either pion daughter from a nominal $\rho$. $B_1(p_{\pi}) = \sqrt{\frac{1+(3p_0)^2}{1+(3p_{\pi})^2}}$ is the Blatt-Weisskopf form factor, as described in Ref.~\cite{BWFF}. \\
The PDFs for \Fsb\ for all components (signal) are sums of two (three)
bifurcated Gaussian functions in each $r$-bin. 

\begin{sidewaystable}
 {\small
\centering
\caption{Summary of the parametrization for each component in the fit model. The details are provided in Ref.~\cite{pit_phd}. Up to three different reconstruction categories are considered for entries with superscript $nT\in 2T,\;1T,\;0T$ (see text). The variables in the subscript list the identifed correlations. The PDFs for the dipion-masses and helicity angles are usually correlated; here we only show a one-dimensional simplification. C stands for sums of Chebyshev polynomials, H for histograms, A for an ARGUS function, (t/db)G for a (triple/double-bifurcated) Gaussian function (the subscript $l$ indicating separate descriptions in each $r$-bin), BW for a relativistic Breit-Wigner function and ${\rm exp}_{\cal R}$ for an exponential function convoluted with a resolution function (${\cal R}$), where the superscript $CP$ labels an included $CP$ asymmetry term depending on the flavor $q$ of $B^{0}_{\rm tag}$. For the signal $\Dt$ PDFs, $|^{\rm truth}_{\rm eff.}$ implies shared $CP$ violation parameters with the respective truth model for longitudinal or transverse polarization, respectively, using an effective lifetime (see text).}
   \begin{tabular}
      {@{\hspace{0.25cm}}c@{\hspace{0.25cm}} @{\hspace{0.25cm}}c@{\hspace{0.25cm}} @{\hspace{0.25cm}}c@{\hspace{0.25cm}} @{\hspace{0.25cm}}c@{\hspace{0.25cm}} @{\hspace{0.25cm}}c@{\hspace{0.25cm}} @{\hspace{0.25cm}}c@{\hspace{0.25cm}} @{\hspace{0.25cm}}c@{\hspace{0.25cm}}  @{\hspace{0.25cm}}c@{\hspace{0.25cm}}}
      \hline \hline
 Mode & \De\ & \Mbc\ & $m(\pi^{\pm}\piz)$ & $\cH^{\pm}$ & \fevt & \Dt \\
   \hline \hline 
Truth (LP) & (dbG+C) & dbG$|_{\De}$ & BW$|_{\De}$ & \Cref{e_helicity}$|_{\De}$ & tbG$_{l}$ & ${\rm exp}_{\cal R}^{CP}$ \\ 
2T (LP) & (dbG+C)$|_{\cH}$ & (dbG+A)$|_{\De\cH}$  & (BW+C)$|_{\cH}$ & H  & dbG$_{l}^{\rm sig}|_{\De}$ & {\rm truth (LP) } \\ 
1T (LP) & (dbG+C)$|_{\cH}$  & (dbG+A) & (BW+C)$^{1T}$  & H$^{1T}$ & dbG$_{l}^{\rm sig}$ & ${\rm exp}_{\cal R}^{CP}|^{\rm truth}_{\rm eff.}$ \\ 
0T (LP) & dbG+C & dbG+A &  BW+C & H &  dbG$_{l}^{\rm sig}$ &   ${\rm exp}_{\cal R}$\\
\hline
Truth (TP) & (dbG+C) & dbG & BW$|_{\De}$   & \Cref{e_helicity} &  dbG$_{l}^{\rm sig}$& ${\rm exp}_{\cal R}^{CP}$ \\ 
2T (TP) &  (dbG+C)$|_{\cH}$ & dbG+A & H  & H & dbG$_{l}^{\rm sig}$& {\rm truth (TP) } \\ 
1T (TP) &  C & dbG+A & H & H & dbG$_{l}^{\rm sig}$ &  ${\rm exp}_{\cal R}^{CP}|^{\rm truth}_{\rm eff.}$ \\ 
\hline 
 $a_1^{\pm}\pi^{\mp}$ & dbG+C & dbG+A & (BW+C)$^{nT}$ & H$^{nT}$ & dbG$_{l}^{\rm sig}$& ${\rm exp}_{\cal R}^{CP}$ \\
 $a_1^{0}[\rho^\pm\pi^{\mp}]\pi^{0}$ & dbG+C & dbG+A & H$^{nT}$ &H$^{nT}$ & dbG$_{l}^{\rm sig}$&${\rm exp}_{\cal R}^{CP}$  \\
 $a_1^{0}[\rho^0\pi^{0}]\pi^{0}$ & dbG+C & dbG+A & C$|_{\cH}$ & G+C & dbG$_{l}^{\rm sig}$& ${\rm exp}_{\cal R}^{CP}$ \\
 $\omega\piz$ & dbG+C & dbG+A & (G+C)$|_{\cH}$ & (G+C) & dbG$_{l}^{\rm sig}$& ${\rm exp}_{\cal R}^{CP}$ \\
$(4\pi)_{\rm NR}$ & dbG+C & dbG+A & C & C &dbG$_{l}^{\rm sig}$ & ${\rm exp}_{\cal R}^{CP}$ \\
 $(\rho^{\pm}\pi^{\mp}\piz)_{\rm NR}$ &(dbG+C)$^{nT}|_{\Mbc,\cH}$ &(dbG+A)$_{\cH}$ & (BW+C)$^{nT}|_{\cH}$ & H$^{nT}$ & dbG$_{l}^{\rm sig}$& ${\rm exp}_{\cal R}^{CP}$ \\
 $(\rho^{0}\piz\piz)_{\rm NR}$ & dbG+C & dbG+A & C & H & dbG$_{l}^{\rm sig}$& ${\rm exp}_{\cal R}^{CP}$ \\
 $(f_{0}\piz\piz)_{\rm NR}$ & dbG+C & dbG+A & C &H & dbG$_{l}^{\rm sig}$& ${\rm exp}_{\cal R}^{CP}$ \\
\hline
continuum  & $C^{nT}|_{\fevt}$ & A & (BW+C)$^{nT}$ & $C^{nT}|_{m(\pi^{\pm}\piz)}$  &dbG$_{l}$ & ${\rm exp}_{\cal R}$\\
$B^0_{\rm charm}$ & C$^{nT}$ & A & (BW+C)$^{nT}|_{\cH}$ & $C^{nT}$ & dbG$_{l}^{\rm sig}$& ${\rm exp}_{\cal R}$ \\ 
$B^+_{\rm charm}$ &  C$^{nT}$ & A & (BW+C)$^{nT}|_{\cH}$ & $C^{nT}$ & dbG$_{l}^{\rm sig}$& ${\rm exp}_{\cal R}$ \\ 
$B^0_{\rm charmless}$ &  C$^{nT}|_{\cH}$ & (dbG+A)$|_{\De}$ & (BW+C)$^{nT}|_{\cH,\De}$ & $C^{nT}|_{\De}$ & dbG$_{l}^{\rm sig}$& ${\rm exp}_{\cal R}$ \\ 
$B^+_{\rm charmless}$ &  C$^{nT}|_{\cH}$ & A & (BW+C)$^{nT}|_{\cH}$ & $C^{nT}$ & dbG$_{l}^{\rm sig}$& ${\rm exp}_{\cal R}$ \\ 
   \hline \hline
   \end{tabular}
\label{t_model_suma}
}
\end{sidewaystable}

\subsection{Signal Model}
\label{signal model}
 The \Brprm\ model consists of seven parts
 (see~\Cref{tab_recEff,t_model_suma}) and is determined from fully
 simulated signal MC events for each $\rho$ polarization state (LP and
 TP) and each reconstruction category. 
For each polarization, the $CP$ violation parameters are made common among
the Truth, 2T and 1T (but not 0T) components. The correlation matrices
for all signal components with longitudinal polarization are given in~\Cref{appendix}.

\subsubsection{Truth Model}
For both polarizations, the truths model's \De\ distributions are described by the sum of two
 bifurcated Gaussians (dbG) and a straight line. The \Mbc\
 PDFs are taken to be dbGs, where 
for longitudinal polarization, the mean and width of the core Gaussian of \Mbc\ depend on $\De$.
\fevt\ is modeled as described above, for transverse polarization, the second and third Gaussians of the
\fevt\ PDFs are taken from longitudinal polarization. 

The \rpm\ mass is modeled with a BW, whose mean and the width depend
slightly on \De. In the fit to data the nominal mean and width of the BW are fixed to the values given
in Ref.~\cite{PDG}. The
PDFs are weighted with a mass-dependent
reconstruction efficiency, being obtained from fully simulated MC
events for each polarization. 

We use the corresponding part of Eq.~\ref{e_helicity} to describe the
helicity distribution of each polarization. The PDFs are weighted
with binned, two-dimensional, symmetrized
($\cH^+\leftrightarrow\cH^-$), helicity-angle dependent reconstruction
efficiencies obtained from fully simulated signal MC events. For
longitudinal polarization we account for the correlation of the
helicity angles with \De\ by using separate reconstruction efficiency
histograms in five bins of \De. 

Each polarizations PDF for \Dt\ is taken to be
\begin{align}
{\cal P}_{{\rm truth},l,s}^{\rm LP(TP)}(\Dt, q)  \equiv& \frac{e^{-|\Delta t|/\tau_{B^0}}}{4\tau_{B^0}}\Bigl\{1-q\Delta w^{l,s} + q(1-2w^{l,s}) \times \nonumber \\
&  \Bigl[{\cal A}_{CP}^{\rm LP(TP)}\cos(\Delta m \Delta t) + {\cal S}_{CP}^{\rm LP(TP)}\sin(\Delta m \Delta t)\Bigr]\Bigr\} \otimes {\cal R}^{s}_{B^0\bar{B}^0}(\Delta t).
\label{e_siglp_dt}
\end{align}
where $w^{l,s}$ accounts for the $CP$ dilution due to the probability
of tagging the wrong $\Btag$ flavor $q$ and $\Delta w^{l,s}$ accounts
for the wrong tag difference between $B$ and $\bar{B}$. Both are
determined from flavor specific control samples. The $B^0$ lifetime,
$\tau_{B^0}$, and the mass difference between
the two mass eigenstates $B^0_H$ and $B^0_L$, $\Delta
m$, are taken from Ref.~\cite{PDG}.
The \Dt\ PDF is convolved with the resolution function described in
\cite{ResFunc}.

\subsubsection{Two Tracks (2T) Model}
 The \De\ distribution is modeled with the sum of a dbG and a straight
 line, where the mean of the core Gaussian, the fraction of the line
 and its slope depend on the helicity angles.
\Mbc\ is described by the sum of a dbG and an Argus function. For LP, their
relative fraction depends on \De\ and the helicity angles.
The \fevt\ distributions are described similar to the ones used for the transverse polarized truth model.
In addition, for longitudinal polarization the widths of each core Gaussians of \fevt\ 
depend on \De.
A wrongly assigned $\pi^0$ is
broaden up the resonance peak in the mass distribution and is also
shifting the helicity angles towards negative values. The
mass PDF is the product of the sum of a second order Chebychev
polynomial and a BW for each mass, where the
relative fractions as well as the width of one of the BWs depends on the
helicity angles. 
A two-dimensional histogram is taken for transverse polarization.
The $\cH^+$-$\cH^-$ PDFs for each polarization are taken to be
two-dimensional histograms. Since only charged tracks contribute to the determination of \Dt, its PDF is identical to the one used in the truth model (see~\cref{e_siglp_dt}).

\subsubsection{One Track (1T) Model}
\label{model1T}
The \De\ distribution is modeled with
the sum of a dbG and a straight line for longitudinal polarization,
where the relative fraction and the slope depend on the helicity angles.
The \De\ distribution for transverse polarization is described by a straight line.
\Mbc\ is described by the sum of a dbG and an Argus function.
The distributions of the dipion masses and helicity angles depend on the charge of the correctly reconstructed $\pi^{\pm}$. 
For longitudinal polarization, the \mppz\ distribution including the
correctly reconstructed $\pi^{\pm}$ is described by the sum of a
BW and a second order Chebychev polynomial, ${\cal
  P}^{\rm OK}(m(\pi^\pm\piz))$, while the \mppz\ distribution with the fake track
from $B_{\rm tag}$ is modeled by the sum of Chebychev polynomials up
to the fifth order, ${\cal P}^{\rm fake}(m(\pi^\pm\piz))$.
The helicity PDFs are taken to be two-dimensional histograms, 
${\cal P}^{\pm}(\cH^{+}, \cH^{-})$, where we
distinguish the two cases of the charge of the correctly reconstructed $\rho^{\pm}$.
The PDF of the $\mpz$-$\mmz$-$\cH^+$-$\cH^-$ distribution is then given by 
\begin{eqnarray}
\begin{aligned}                                                                                                                                     
  & {\cal P}_{\rm 1T}^{\rm LP}(\mpz, \mmz, \cH^{+}, \cH^{-}) \equiv \nonumber \\ 
 & f_{+}{\cal P}^{\rm OK}(\mpz){\cal P}^{\rm fake}(\mmz ){\cal P}^{+}(\cH^{+}, \cH^{-})\; + \nonumber \\
& (1-f_{+}){\cal P}^{\rm fake}(\mpz ){\cal P}^{\rm OK}(\mmz){\cal P}^{-}(\cH^{+}, \cH^{-} ) 
\label{e_1cPiMH}                                                                                                                                   
\end{aligned}
\end{eqnarray} 
where the fraction of events with a correctly reconstructed \pip, $f_{+}$ is made common among the detector configurations SVD1 and SVD2. We ignore such a correlation for transverse polarization, since $f_L$ has been measured to be close to
one~\cite{Brprm_Belle,rhorho_Belle,rhorho_BABAR}. 
The $\rho$ masses and helicity angle distributions for transverse
polarization are each modeled with two-dimensional histograms.

The PDF for \fevt\ is similar to the one used for the transverse polarized truth model.
Even with only one correctly reconstructed track it is possible to use
the $\Delta t$ distribution to obtain $CP$ violation related
information. The \Dt\ PDF is described by~\Cref{e_siglp_dt} with an effective lifetime that accounts for the contamination from the wrongly assigned track.

\subsubsection{Zero Track (0T) Model}

Because the transverse polarization's fraction without any correctly
reconstructed tracks is negligible, we include those events in the model used for transverse polarization when one $\pi^{\pm}$ is correctly reconstructed.
For longitudinal polarization, the \De\ distribution is modeled with the sum of a dbG and a first order Chebychev polynomial
and the \Mbc\ distribution is described by the sum of a dbG and an Argus function.
Each \mppz\ distribution is modeled with the sum of a BW, a second and a third order Chebychev polynomial
and the PDF for the $\cH^+$-$\cH^-$ distribution is taken to be a histogram.
The \Dt\ PDFs for the ``0T'' components are parameterized as
\begin{equation}                                                                                                                                      
{\cal P}_{\rm 0T}^{\rm LP(TP)}(\Delta t) \equiv \frac{1}{2\tau_{\rm 0T}}e^{-|\Delta t|/\tau_{\rm 0T}} \otimes {\cal R}_{\rm 0T}(\Delta t), 
   \label{e_qqdt}      
\end{equation}         
with an effective lifetime $\tau_{\rm 0T}$ and convoluted with the sum of two Gaussian functions ($G(x,\mu,\sigma)$) with a common mean $\mu$                                                                                      
 \begin{equation}                                                                                                                                     
  {\cal R}_{\rm 0T}(\Delta t) \equiv (1-f_{\rm tail})G(\Delta t, \mu, S_{\rm core}\sigma) + f_{\rm tail}G(\Delta t, \mu, S_{\rm core}S_{\rm tail}\sigma).           
\label{q_resqq}                                                                                                                                        
\end{equation}                                                                                                                                         
The second Gaussian function accounts for a broader tail and its width is related to that of the core Gaussian $S_{\rm core}\sigma$ through a multiplicative factor $S_{\rm tail}$.
The scale factor $S_{\rm core}\equiv \sqrt{\sigma_{\rm Rec}^2 + \sigma_{\rm tag}^2}/\beta\gamma c$ is an event-dependent error on \Dt\ constructed from the vertex resolution of $B^{0}_{CP}$ ($\sigma_{\rm Rec}$) and $B^{0}_{\rm tag}$ ($\sigma_{\rm tag}$). We use a different set of Gaussian functions if at least one of the $B$ vertices is obtained from only one track.

\subsection{Continuum Model}
\label{qq model}
 The continuum model consists of three components ($2T$, $1T$ and $0T$) and is studied with continuum MC simulation. 
To reduce the systematic uncertainty related to a fixed parametrization, some parameters of the continuum model are floated in the fit to data; the initial values for these model parameters are obtained from the fits to MC simulation. 
The PDF description is confirmed with on-resonance (almost entirely continuum events, after reconstruction) and off-resonance data. The correlations between the helicity angles and the dipion masses are taken into account and this description is confirmed by projecting into several slices of one of the variables and comparing the projected shapes of the other distribution in the two data sets.\\

The \De\ distributions of all continuum components are described by the sum of a first and a second order Chebychev polynomial, 
where the first order one depends on \fevt. The dependence slightly
differs for the three different reconstruction categories. The \Mbc\ distributions of all reconstruction types is commonly described by an Argus function.
 Because we find a small difference in the $r$-bin distributions of the $\pm,0$ and $2c\pi$ components, we use different $r$-bin fractions for each of the three categories.

For the description of the mass and helicity angle distributions we
use certain combination of two kinds of one-dimensional PDFs; one for
distributions including a $\rho^{\pm}$ resonance and one otherwise.
The different reconstruction types are then described by combinations
of these PDFs. This reduces the degrees of freedom in the fit to data,
where the parameters of the continuum model are floated. The mass PDF
for the distributions including a $\rho^{\pm}$ resonance is given by a
sum of a BW and Chebychev polynomials. The sum of a bifurcated Gaussian and Chebychev polynomials is describing the combinatorial background.                                                                                                                    
The helicity PDFs for for distributions for both cases, including a
$\rho^{\pm}$ resonance and otherwise, are given by the sums of Chebychev
polynomials up to the ninth order. Due to a correlation with the dipion
masses, some of the helicity parameters depend on the dipion masses.

The \Dt\ PDF is the sum of ~\Cref{e_qqdt} and a delta function. The latter accounts for the prompt production of light quarks ($u,d,s$) in addition to the exponential decay that describes the production of charm quarks with an effective lifetime $\tau_{q\bar{q}}$: 
\begin{equation}                                                                                                                                      
{\cal P}_{q\bar{q}}(\Delta t) \equiv \bigl[(1-f_{\delta})\frac{e^{-|\Delta t|/\tau_{q\bar{q}}}}{2\tau_{q\bar{q}}} + f_{\delta}\delta(\Delta t - \mu_{q\bar{q}})\bigr] \otimes {\cal R}_{q\bar{q}}(\Delta t).
   \label{e_ofres_dt}      
\end{equation}       
The resolution function ${\cal R}_{q\bar{q}}$ is similar
to~\Cref{q_resqq}.

\subsection{\BBbar\ Model}
\label{BB model}
The model for combinatorial background from other $B$ decays is obtained from four separate sets of MC simulation: neutral and charged $B$ meson decays into charmed and charmless final states. The samples contain $10$ and $50$ times the number of expected charmed and charmless events in the data, respectively. 
In the case of neutral $B$ meson decays into charmless final states,
the four-pion modes are excluded from the MC sample as they are
treated separately (see~\Cref{B4pi model}). In the fit model, we
further distinguish the reconstruction categories $2T$, $1T$ and $0T$;
the $2T$ category is only a significant contribution for charged $B$
decays. The fraction of events for each charged $B$ model is fixed
relative to the number of the corresponding neutral model as obtained
from MC simulation and is given in~\Cref{tab_bf_fixed} (the
distributions for charm final states being almost identical). The \Dt\
PDFs are similar to~\Cref{e_qqdt}, with an effective lifetime for each
component.

\subsubsection{\texorpdfstring{\mbox Charm {\boldmath $B^0$} Backgrounds}{Charm Neutral B Backgrounds}}
\label{model generic}

 The \De\ distributions for all three reconstruction categories of charm $B^0$ decays are described by the sum of Chebychev polynomials up to the second order, and the \Mbc\ distributions are
described by Argus functions. The PDF of the $m(\pi^\pm\piz)$
distributions including one correctly reconstructed $\rho^{\pm}$ meson
is given by the sum of a BW, a second and a third order Chebychev polynomial, where the fraction of the BW depends on the helicity angles.
The distributions without a correctly reconstructed $\rho^\pm$ meson
are take to be sums of Chebychev polynomial up to the fifth order.
The distributions of the helicity angles are modeled by the products
of sums of Chebychev polynomials up to the eighth order for each
reconstruction category.

\subsubsection{\texorpdfstring{\mbox Charm {\boldmath $B^\pm$} Backgrounds}{Charm Charged B Backgrounds}}
\label{model gc}


The PDFs of the \De, \Mbc, \mppz\ and helicity distributions are similarly described as those
of neutral $B$ decays into charm final states. For
the mass PDF, a
correlation with the helicity angles is included for
the component including a correctly reconstructed $\rho^\pm$ meson.
 For the 2T component, the mass and
 helicity distributions are taken to be the sum of a BW and a
 second order Chebychev polynomial and the sums of
 Chebychev polynomials up to the eighth order, respectively.

\subsubsection{\texorpdfstring{\mbox Charmless {\boldmath $B^0$} Backgrounds}{Charmless Neutral B Backgrounds}}
\label{model rm}

The \De\ distributions are described by sums of Chebychev
polynomials up to the forth order, where a correlation with the
helicity angles is accounted for in the case of a correctly
reconstructed $\rho^\pm$ meson.
 The \Mbc\ distributions are described by Argus functions. In the
 case of a correctly
reconstructed $rho^\pm$ resonance, a dbG is added. Its relative fraction 
 depends on \De\ and the helicity angles.

The PDF of the \mpz-\mmz\ distribution of the 1T component
is given by the the product of a BW added to the sum of
Chebychev polynomials up to the third order for each dipion mass,
where the fraction of the BW of the correctly reconstructed
resonance depends on the helicity angles.
The product of a sum of Chebychev polynomial up to the fifth order is taken if no $\rho^{\pm}$ resonance has been correctly reconstructed.
The distribution of the helicity angles is modeled by the product of
sums of Chebychev polynomials up to the eighth order for each
reconstruction category.
A correlation with \De\ is accounted for in the case of a correctly
reconstructed $\rho^\pm$ resonance.

\subsubsection{\texorpdfstring{\mbox Charmless {\boldmath $B^\pm$} Backgrounds}{Charmless Charged B Backgrounds}}
\label{RC}
The $\De$ distributions are described by sums of Chebychev polynomials
up to the third order. A correlation with the helicity angles is
included for the reconstruction category 1T.
The PDF for $\Mbc$ is an Argus function. The mass distributions are
 described by the sum of a BW and Chebychev polynomials in the case of
 a correctly reconstructed $\rho^\pm$ resonance and by sums of Chebychev
 polynomials otherwise.
A correlation of the $\rho^\pm$ masses with the helicity angles is included in the description
of all reconstruction categories. The PDFs for the helicity angles are
taken to be sums of Chebychev polynomials.

\subsection{Peaking Background Model}
\label{B4pi model}
 The PDFs of the remaining four-pion states are determined from
 individually generated MC samples. We consider the following final
 states from $B^0$ decays (the subscript NR denoting a non-resonant
 multi-meson state): $(\pip\piz\pim\piz)_{\rm NR}$,
 $a_1^{\pm}[\rho^\pm\piz]\pi^{\mp}$,  $a_1^{0}\pi^{0}$,
 $(\rho^\pm\pi^\mp\piz)_{\rm NR}$, $(\rz\piz\piz)_{\rm NR}$,
 $(f_0(980)[\pip\pim]\piz\piz)_{\rm NR}$ and
 $\omega[\pip\pim\piz]\piz$. The numbers of expected $B\to
 a_1^\pm\pi^\mp$ and $B\to \omega\piz$ events are fixed according to
 their world average branching fractions~\cite{PDG}; the other
 four-pion modes are poorly known and their yields are allowed to
 float in the fit. \\
Since only one dipion combination from $B\to \omega[\pip\pim\piz]\piz$
lies in the signal window, this non-peaking mode has a very small
reconstruction efficiency: only one event is expected and the model is
described in Ref.~\cite{pit_phd}.\\ 
If no other description is explicitly mentioned, we use a dbG to model the \De\ distributions of all peaking backgrounds. The tail Gaussian is obtained from correctly reconstructed \Brprm\ MC events and is made common among all four-pion final states. The sum of Chebychev polynomials up to the third order is added in order to describe underlying combinatorial background from wrongly assigned tracks.
The \Mbc\ distributions of the four-pion final states are described by
a dbG for correctly reconstructed tracks plus an Argus function for
the combinatorial background. The \fevt\ distribution of all four-pion states are described similarly
to the one used for the transverse polarized truth model and each \Dt\
PDF is similar to~\Cref{e_siglp_dt}.

\subsubsection{ \texorpdfstring{\mbox Model For {\boldmath $B^0\to
      (\pip\piz\pim\piz)_{\rm NR}$} Decays}{Model For B->pi+pi0pi-pi0 Decays}}  
\label{Bto4pimodel}

The distribution of each mass is modeled by the sum of a second and a third order Chebychev polynomial  
and the distribution of each helicity angle is described by the sum of
Chebychev polynomials up to the eighth order.

\subsubsection{ \texorpdfstring{\mbox Model For {\boldmath $B^0\to a_1\pi$} Decays}{Model For B->a1pi Decays}}  
\label{a1pimodel}

 We consider both decays $B^0\to a_1^{\pm}\pi^{\mp}$ and $B^0\to a_1^{0}\pi^{0}$ separately, where the latter decay is further separated into two different possible decays of the $a_1^0$: $a_1^0 \to \rho^{\pm}\pi^{\mp}$ and $a_1^0 \to \rho^{0}\pi^{0}$.\\  

\textbullet {\; Model for $B^0\to a_1^{\pm}\pi^{\mp}$ Decays}\\

We consider only the subsequent decay $a_1^{\pm}\to\rho^{\pm}\pi^0$ as
the decay $B^0\to a_1^{\pm}[\rho^0\pi^{\pm}]\pi^{\mp}$ yields in a
different final state and is included in the non-peaking $B^0$ decays
into charmless final states. We assume isospin symmetry and set the
fraction of $a_1^{\pm}$ decaying to $\rho^{\pm}\pi^0$ to be $50\%$.
Because of the high momentum of the $\pi^{\mp}$, the helicity angle associated with a $\pi^\pm\piz$ pair reconstructed with the $\pi^{\mp}$ from the $B$ decay peaks sharply at $\cH=-1$. Therefore, the helicity angles are especially useful in separating this component from others.
We consider a 1T and a 0T contribution in the description of the mass and
helicity PDFs.
The mass PDF for the 1T part is the sum of a BW and a second
order Chebychev polynomial for the correctly reconstructed $\rho^\pm$
meson multiplied with the sum of Chebychev polynomials.
The mass distribution without a correctly reconstructed $\rho$
resonance is described by the sum of Chebychev polynomials up to the
fifth order for each mass. The helicity PDFs for all cases of reconstruction are histograms from
fully simulated MC events.\\

\textbullet {\;Model For $B^0\to a_1^{0}\pi^{0}$ Decays}\\

Three dominant subsequent decays of the $a_1^0\to\rho\pi$ yield
in the same final state as our signal, $a_1^0\to\rop\pim, \rom\pip, \rz\piz$. We assume that their amplitudes are of similar size, because of isospin arguments. 
A common PDF is used to describe the \Dt\ distribution.\\ 

- {\;$B^0\to a_1^{0}[\rho^{\pm}\pi^{\mp}]\pi^{0}$}\\
 We have a common model for $a_1^{0}\to\rho^{\pm}\pi^{\mp}$ decays, where we furthermore distinguish between the different reconstruction categories 1T and 0T, due to misreconstruction.
Opposite to $B^0\to a_1^{\pm}\pi^{\mp}$ decays, the fast pion is neutral, therefore the corresponding helicity distributions peak at $\cH=+1$.
The mass and helicity PDFs for all reconstruction categories are individual two-dimensional histograms.
We use a PDF similar to \cref{e_qqdt} with an effective lifetime to
account for the contamination from wrong side tracks for the reconstruction category 0T. \\

 - {\;Model For $B^0\to a_1^{0}[\rho^{0}\pi^{0}]\pi^{0}$ Decays}\\
This decay does not contain a $\rho^{\pm}$ resonance, hence no
separate treatment is needed. 
The mass PDF is the product of distinct sums of Chebychev
polynomials up to the fifth order, whose combinations depend on the
helicity angles (if $\cH^+>\cH^{-}$ or else).
The PDF for the helicity angles is the product of sums of two Gaussians and a second order Chebychev polynomial

\subsubsection{ \texorpdfstring{\mbox Model For {\boldmath $B^0\to
      (\rho^\pm\pi^\mp\piz)_{\rm NR}$} Decays}{Model For B->rho+pi-pi0 Decays}}  
\label{roppimpi0model}


Since there is no suitable decay model for a pseudo-scalar decaying
into a vector particle and two pseudo-scalars, we assume a phase-space
model and account for that assumption in the systematic uncertainty.
We consider the categories 1T and 0T separately.
For the category 1T a correlation of \De\ with \Mbc\ and the helicity angles is accounted
for by introducing a dependence of the relative fraction of the
dbG of the \De\ PDF.
 The \De\ distribution for reconstruction category 0T is described by a first order Chebychev polynomial.
The relative fraction of the dbG of the \Mbc\ PDF for events of category $1T$ 
depends on the helicity angles.
The mass distributions for both reconstruction categories are
described similar to the $B^0\to a_1^\pm\pi^\mp$ model, in
addition, the fraction of the resonant part depends on the helicity angles for the category $1T$.
 The PDF for the helicity distribution is taken to be two-dimensional histogram for each reconstruction category.

\subsubsection{ \texorpdfstring{\mbox Model For {\boldmath $B^0\to
      (\rz\piz\piz)_{\rm NR}$} And {\boldmath $B^0\to (f_0\piz\piz)_{\rm NR}$} Decays}{Model For B->X0pi0pi0 Decays}}  
\label{X0pi0pi0model}  

 The modes $B\to (\rz\piz\piz)_{\rm NR}$ and $B\to (f_0\piz\piz)_{\rm
   NR}$ have almost identical distributions and are therefore combined
 to one component, referred to as $X^0\piz\piz$. These decays exhibit
 certain kinematic behaviors of the dipion masses (flat distribution)
 and the helicity angles. Because the $\piz$ momenta are usually higher then those of the
charged daughters of the $\rho^0\;(f_0)$, the helicity angles peak at $\cH=+1$. This kinematic behavior also occurs in other combinatorial backgrounds, {\it e.g.} other $B$ decays involving a $\rho^0$ meson. Therefore, we add a component to the $X^0\piz\piz$ model where the PDFs of dipion masses and helicity angles are taken from the $X^0\piz\piz$ model, while the PDFs for \De\ and \Mbc\ are taken from a combinatorial background ($B\bar{B}$ model). We determine the fraction of the combinatorial description within the $X^0\piz\piz$ model, $f_{\rm comb}^{X^0\piz\piz}$, in the fit to data in order to avoid a misidentification of combinatorial background as $B\to X^0\piz\piz$ due to the strong discrimination power of the helicity angles in this case.\\
The PDF for the mass distribution is take to be the product of a sum
of a second and third order Chebychev polynomial for each dipion mass and
the helicity PDF is taken to be a two-dimensional histogram.

\subsection{Full Model}
\label{full model}
The total likelihood for $216176$ events in the fit region is
\begin{equation}
  {\cal L} \equiv \prod_{l,s} \frac{e^{-\sum_{j}N^{s}_{j}\sum_{l,s}f^{l,s}_{j}}}{N_{l,s}!} \prod^{N_{l,s}}_{i=1} \sum_{j}N^{s}_{j}f^{l,s}_{j}{\cal P}^{l,s}_{j}(\De^{i},\Mbc^{i},\Fsb^{i},M_+^{i}, M_-^{i}, H_+^{i}, H_-^{i}, \Dt^{i}, q^{i}),
\end{equation}
which runs over event $i$, component $j$, $r$-bin $l$ and SVD configuration $s$. Instead of two free signal yields $N^{s}_{\rm Sig}$ for each detector configuration, the branching fractions for the four-pion final states ($j\in 1,\;...\;7$) are chosen as single free parameters ${\cal B}(B\to X)$ and incorporated into the fit as
\begin{equation}
  N^{s}_{j} = {\cal B}(B^{0}\to f)N^{s}_{\BBbar}\epsilon^{s}_{j}\eta,
\label{e_N}
\end{equation}
where $\epsilon^{s}_{j}$ are the signal selection efficiencies, fixed to the values listed in~\Cref{tab_recEff}.
Using independent control samples, we determine the efficiency correction factor $\eta=\eta^\pm\cdot\eta^0$ that accounts for differences between data and MC in the charged particle identification for the two charged pions, $\eta^\pm=0.93\pm0.03$, and $\piz$ reconstruction for two $\piz$ mesons, $\eta^0=0.91\pm0.03$. The uncertainties on both corrections are included in the systematic uncertainties of our results (see~\Cref{Systematic Uncertainties}). 

\Cref{e_N} takes the distinct forms for the two possible polarization states of the $\rho$ meson: for longitudinally polarized $\rho$ mesons (LP),
\begin{equation}
N_{\rm LP}^{s}  = {\cal B}(\Brprm)f_{L}N_{\BBbar}^{s}\epsilon_{\rm LP}^{s}\eta_{\rm LP}^{s},
\end{equation}
and similarly for transversely polarized $\rho$ mesons, with $(1-f_L)$ replacing $f_L$. The fraction of events in each $r$-bin $l$ for component $j$ is denoted by $f^{l,s}_{j}$ and fixed according to MC simulation for all $B$ decays.

In the fit to data, we float 94 parameters in total. Besides the branching fraction, $f_L$ and the $CP$ violating parameters ${\cal A}_{CP}$ and ${\cal S}_{CP}$ of the decay \Brprm, the free parameters are the branching fractions of $B\to (\pip\piz\pim\piz)_{\rm NR},\; (\rho^\pm\pi^\mp\piz)_{\rm NR},\; a_1^0\piz, X^0\piz\piz$, as well as the yields $N^{s}_{\qqbar}$, $N^{{\rm charm}, s}_{\BzBzb}$ and $N^{{\rm charmless}, s}_{\BzBzb}$. The remaining free parameters describe the shape of the continuum model. The remaining yields are fixed to the values determined from MC simulation as given in~\Cref{tab_bf_fixed,tab_pb}.  

\begin{table}
  \centering
 \caption{Summary of the fixed ratios of the yields of charmed and charmless \BpBm\ background relative to the respective floated number of \BzBzb\ background events for the two detector configurations $s$. The central values are obtained from MC simulation; the errors are statistical.}
   \begin{tabular}
    {@{\hspace{0.5cm}}c@{\hspace{0.25cm}}  @{\hspace{0.25cm}}c@{\hspace{0.25cm}}  @{\hspace{0.25cm}}c@{\hspace{0.5cm}}}
    \hline \hline
    Component & Yield SVD1 & Yield SVD2\\
    \hline
     $N^{{\rm charm},s}_{\BpBm}$ & $(1.78 \pm 0.02)N^{\rm charm, SVD1}_{\BzBzb}$ & $(2.02 \pm 0.01)N^{\rm charm, SVD2}_{\BzBzb}$\\[5pt]
    $N^{{\rm charmless},s}_{\BpBm}$ & $(1.04 \pm 0.02)N^{\rm charmless, SVD1}_{\BzBzb}$ & $(1.00 \pm 0.01)N^{\rm charmless, SVD2}_{\BzBzb}$\\[5pt]
    \hline \hline
  \end{tabular}
  \label{tab_bf_fixed}
\end{table}

For \De, \Mbc\ and \Fsb, we incorporate calibration factors to correct for possible differences between the data and MC distributions. They are determined from a large-statistics control sample \drho\ and are used to calibrate the means and widths of the core bifurcated Gaussian functions for the \De\ and \Mbc\ PDFs of all four-pion final states, and of the \Fsb\ PDFs of all $B\bar{B}$ components. Similarly, we correct the fractions of events in each $r$-bin for all $B\bar{B}$ components. Furthermore, the core Gaussian functions of the signal are contraint to be common for all four-charged-pion final states for \De\ and \Mbc\ and for all $B\bar{B}$ modes for \Fsb.

\begin{table}
  \centering
 \caption{List of peaking backgrounds, assumed branching fractions and their expected yields $N^{\rm s}_{\rm expected}$ for the two detector configurations; SVD1 and SVD2.}
   \begin{tabular}
    {@{\hspace{0.5cm}}c@{\hspace{0.25cm}}  @{\hspace{0.25cm}}c@{\hspace{0.25cm}}  @{\hspace{0.25cm}}c@{\hspace{0.25cm}}  @{\hspace{0.25cm}}c@{\hspace{0.25cm}}  @{\hspace{0.25cm}}c@{\hspace{0.5cm}}}
    \hline \hline
    Mode & ${\cal B}$ $(\times 10^{-6})$ & $N^{\rm SVD1}_{\rm expected}$ & $N^{\rm SVD2}_{\rm expected}$ \\
    \hline
  $\Bz \rightarrow \aone [\pi^{\pm} \pi^{0} \pi^{0}] \pi^{\mp}$ &  $0.5\times (26 \pm 5)$&  $11$ & $52 $\\
 $\Bz \rightarrow \omega[\pi^{+} \pi^{-} \pi^{\pm}] \pi^{0}$ & $0.5\pm0.5$ & $0$ & $1$\\
     \hline \hline
  \end{tabular}
  \label{tab_pb}
\end{table}

\section{Results}
\label{Fit Result}
\begin{figure}[h]
  \centering
  \includegraphics[height=150pt,width=!]{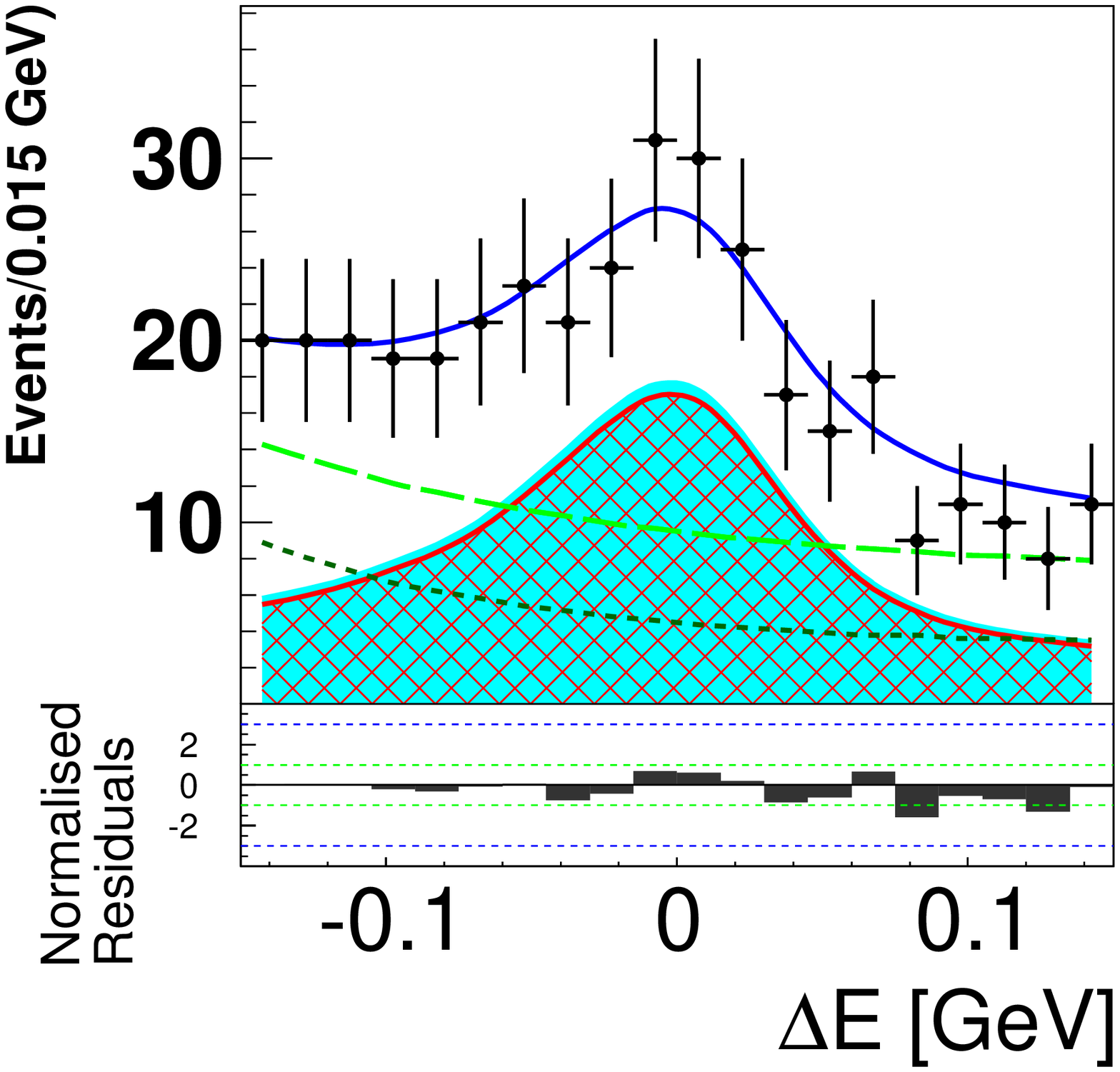}
   \includegraphics[height=150pt,width=!]{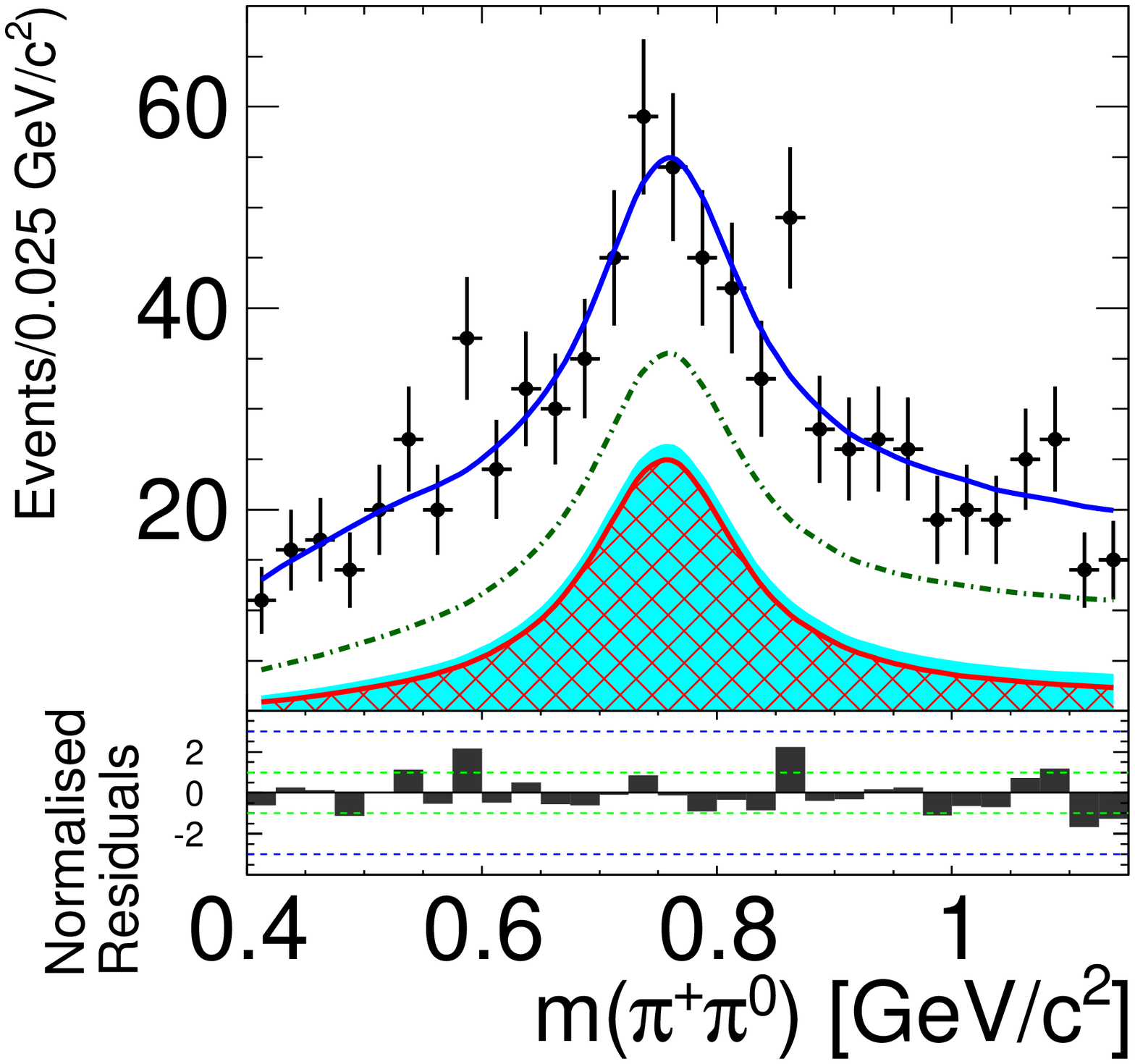}
  \includegraphics[height=150pt,width=!]{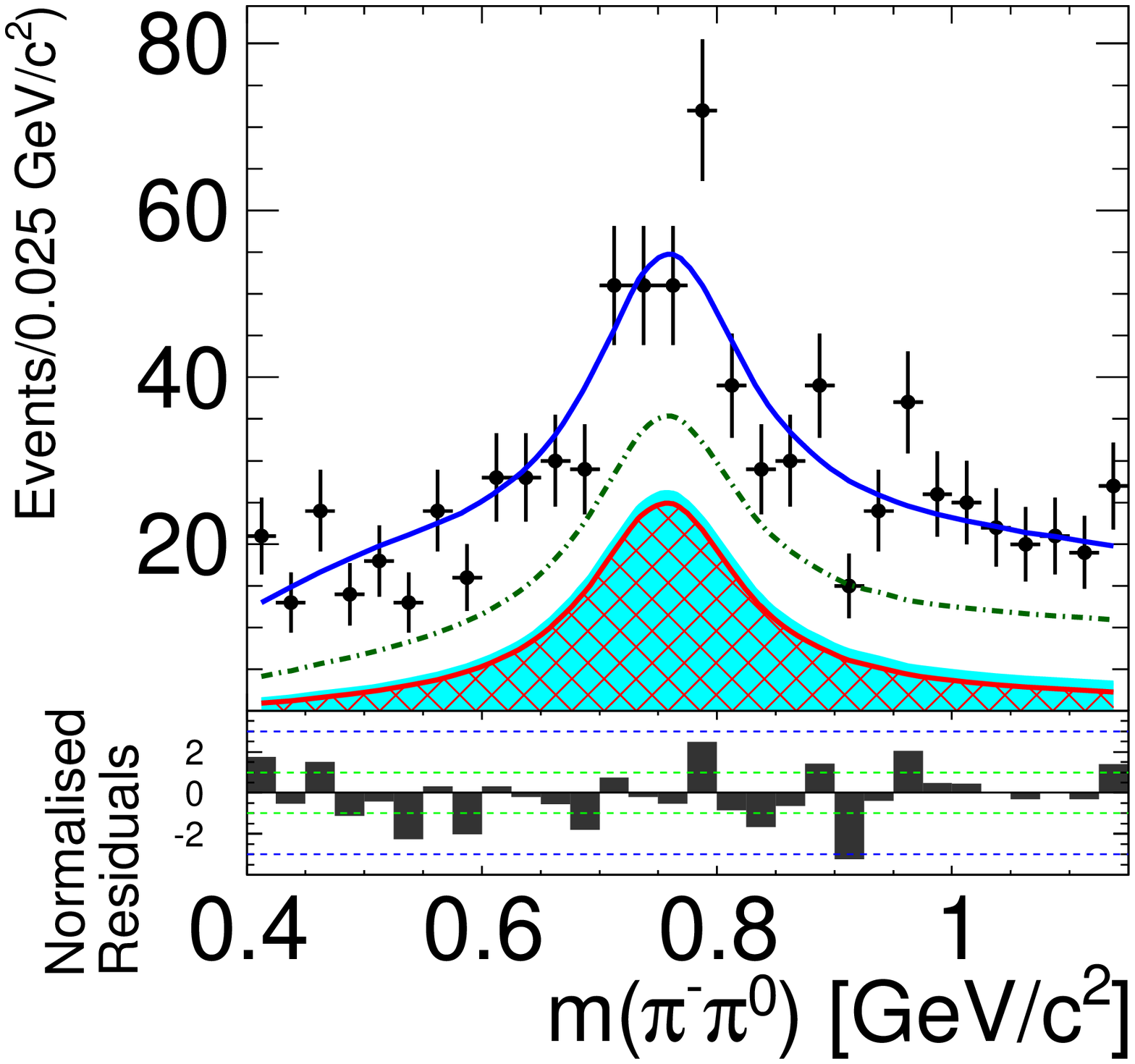}
\put(-350,120){(a)}
  \put(-195,120){(b)}
 \put(-35,120){(c)}

  \includegraphics[height=150pt,width=!]{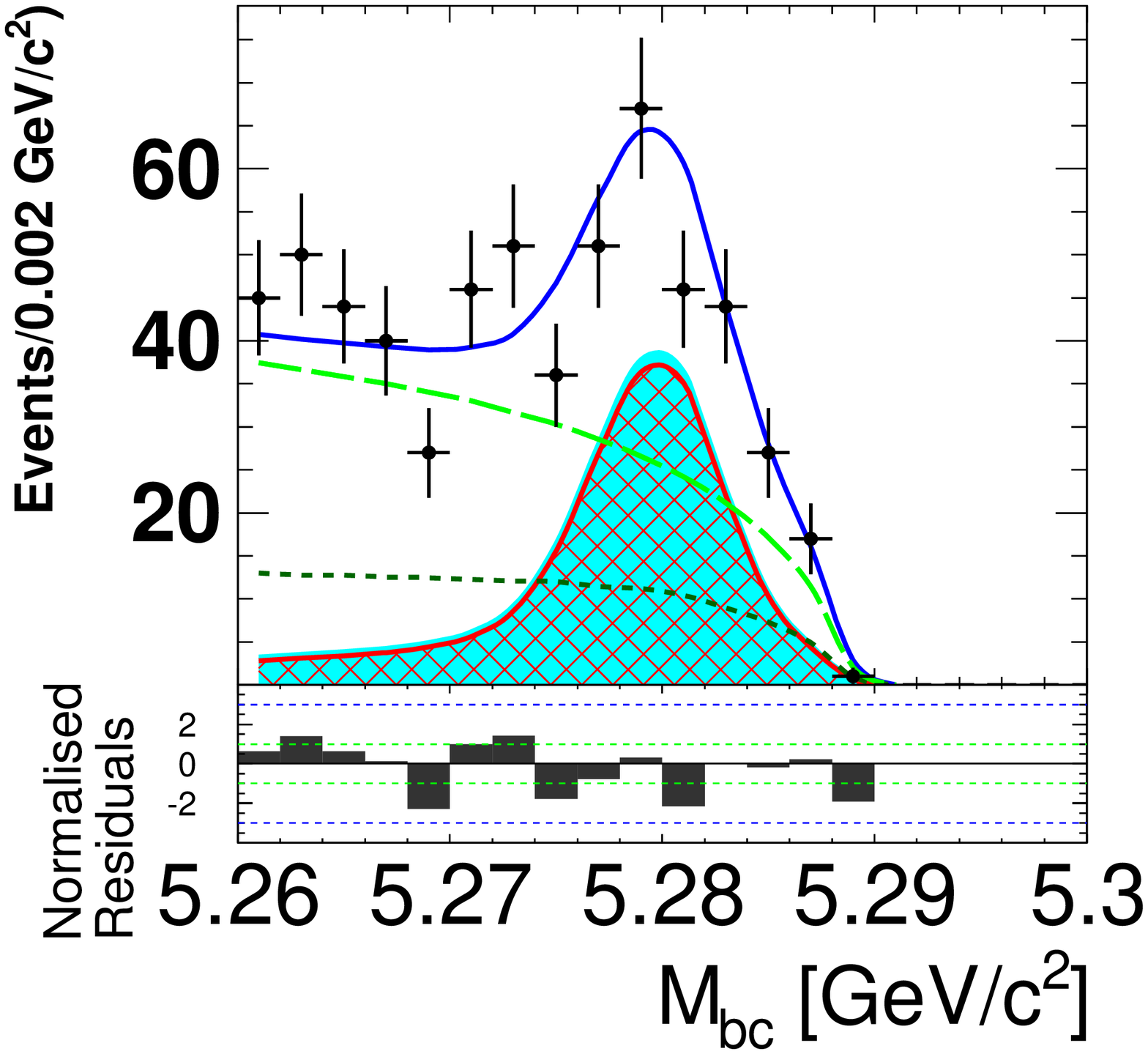}
   \includegraphics[height=150pt,width=!]{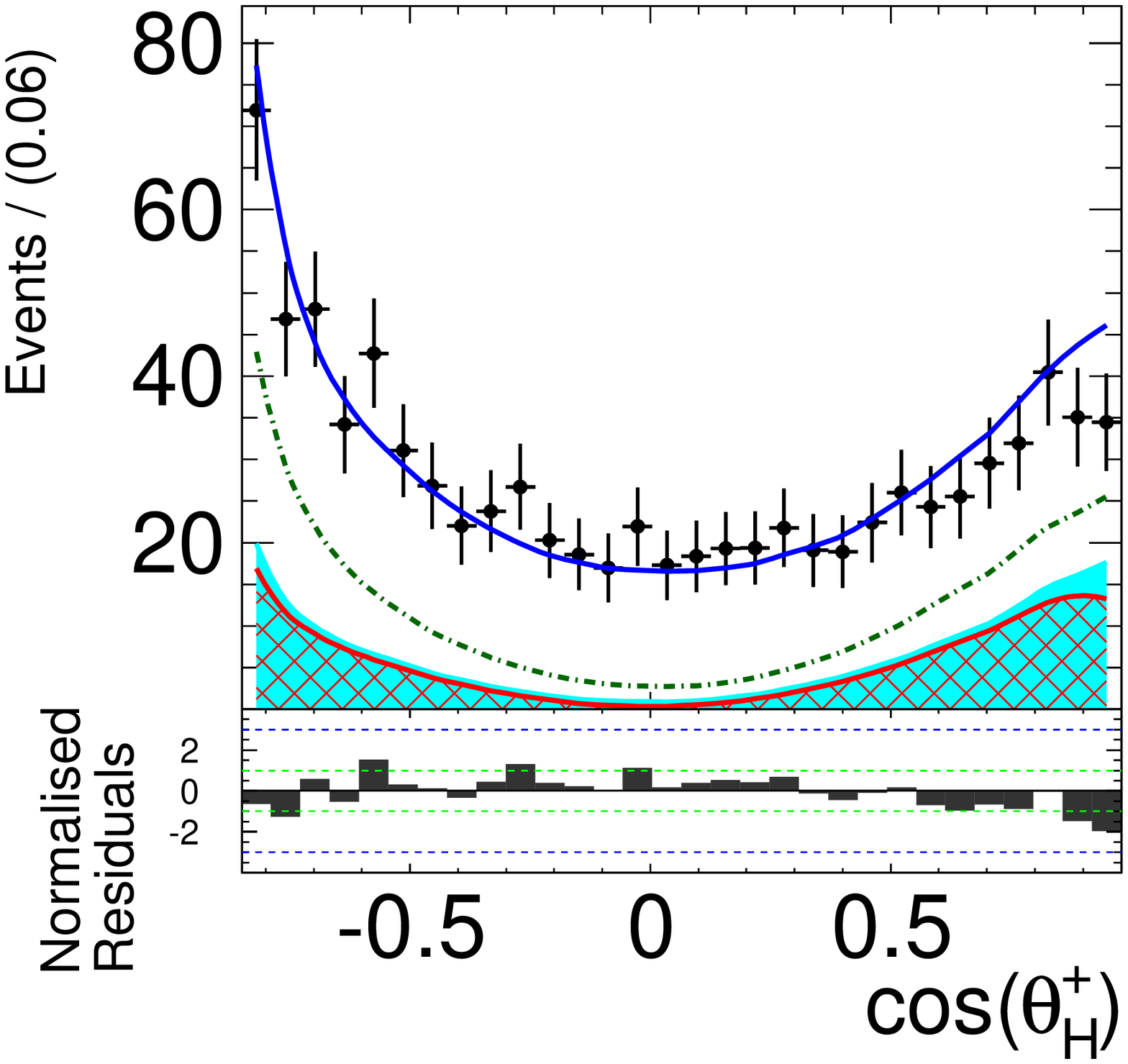}
  \includegraphics[height=150pt,width=!]{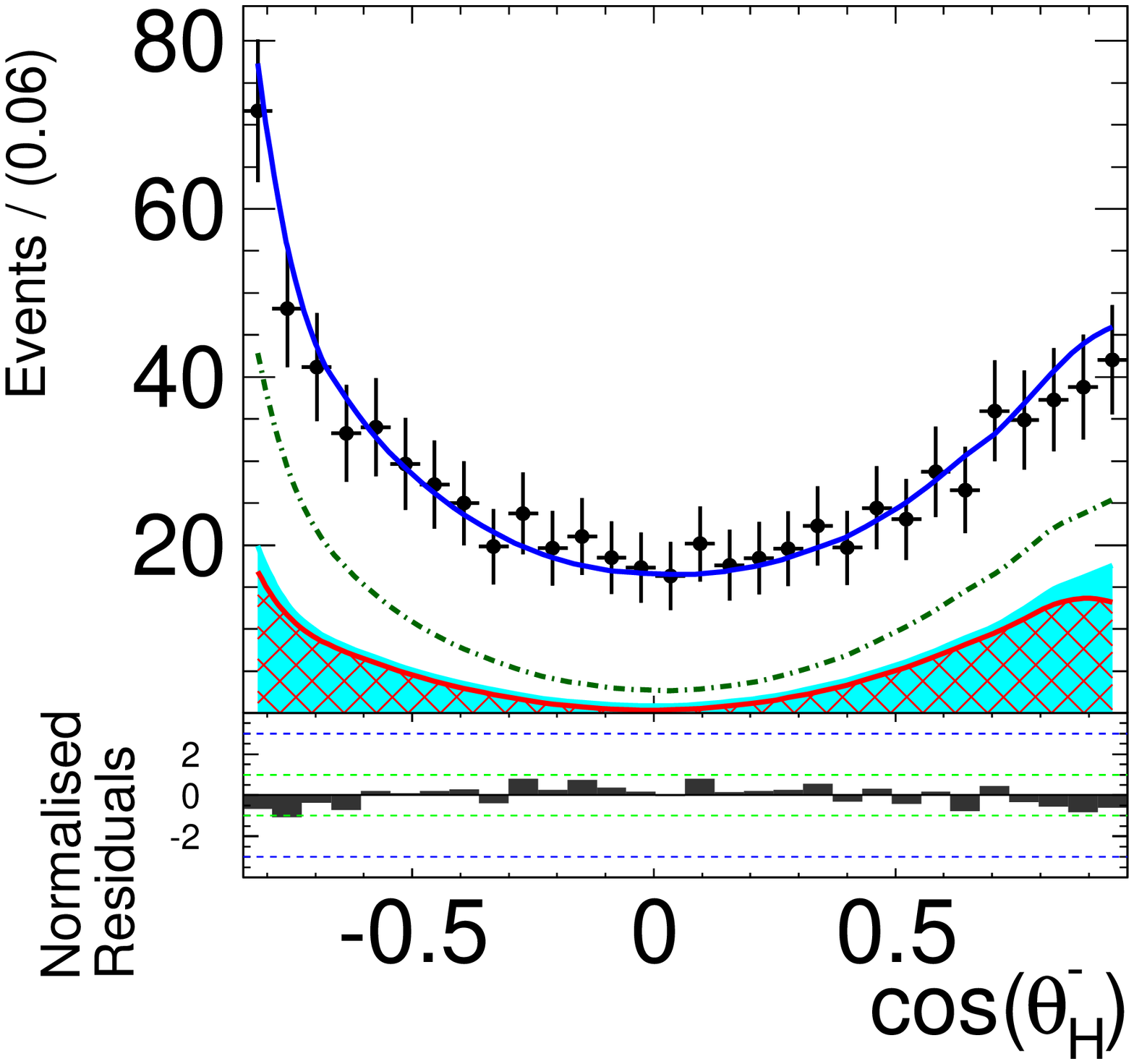}
\put(-350,120){(d)}
  \put(-195,120){(e)}
 \put(-35,120){(f)}\\
\includegraphics[height=150pt,width=!]{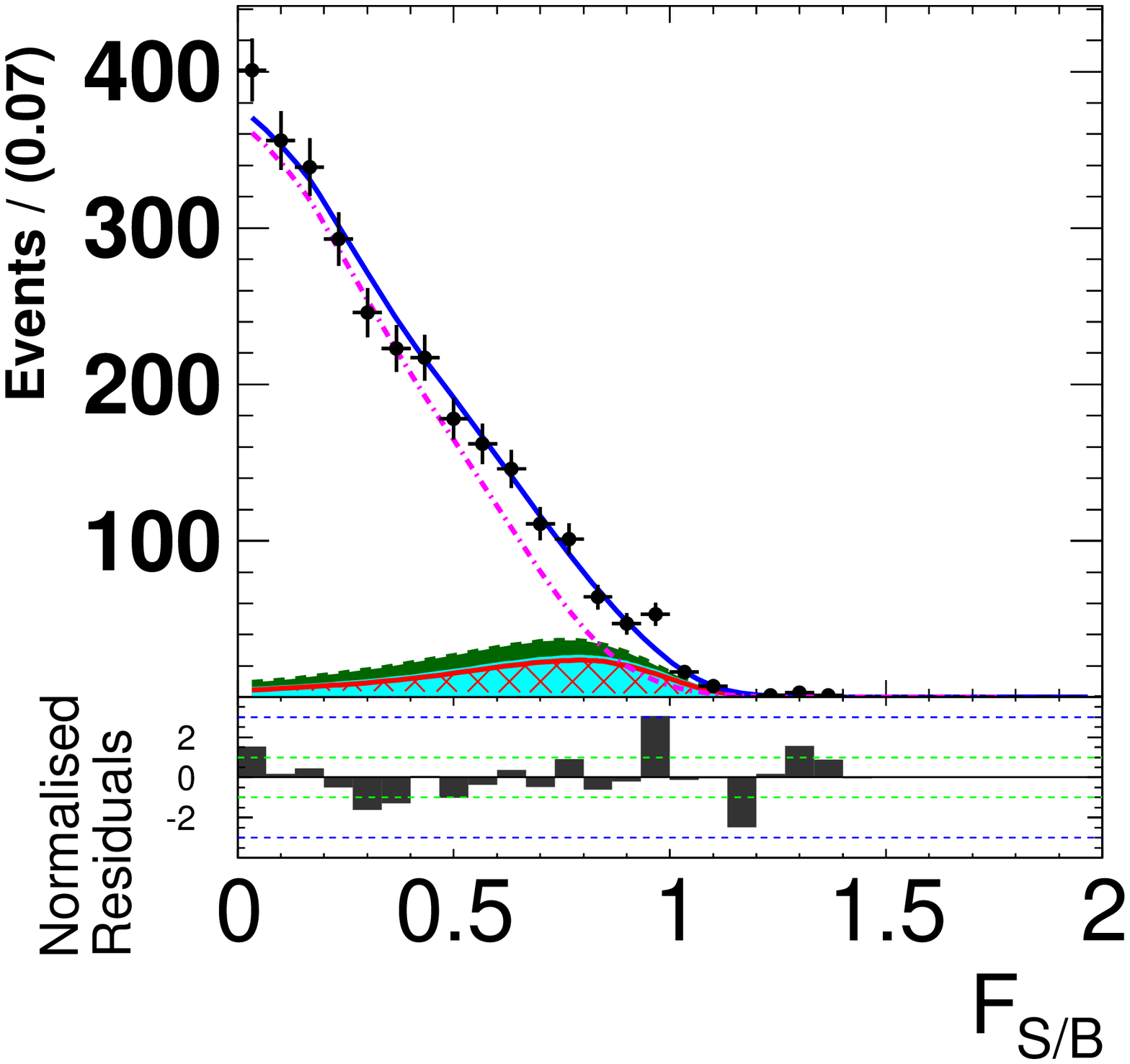}
 \put(-35,120){(g)}
 \includegraphics[height=150pt,width=!]{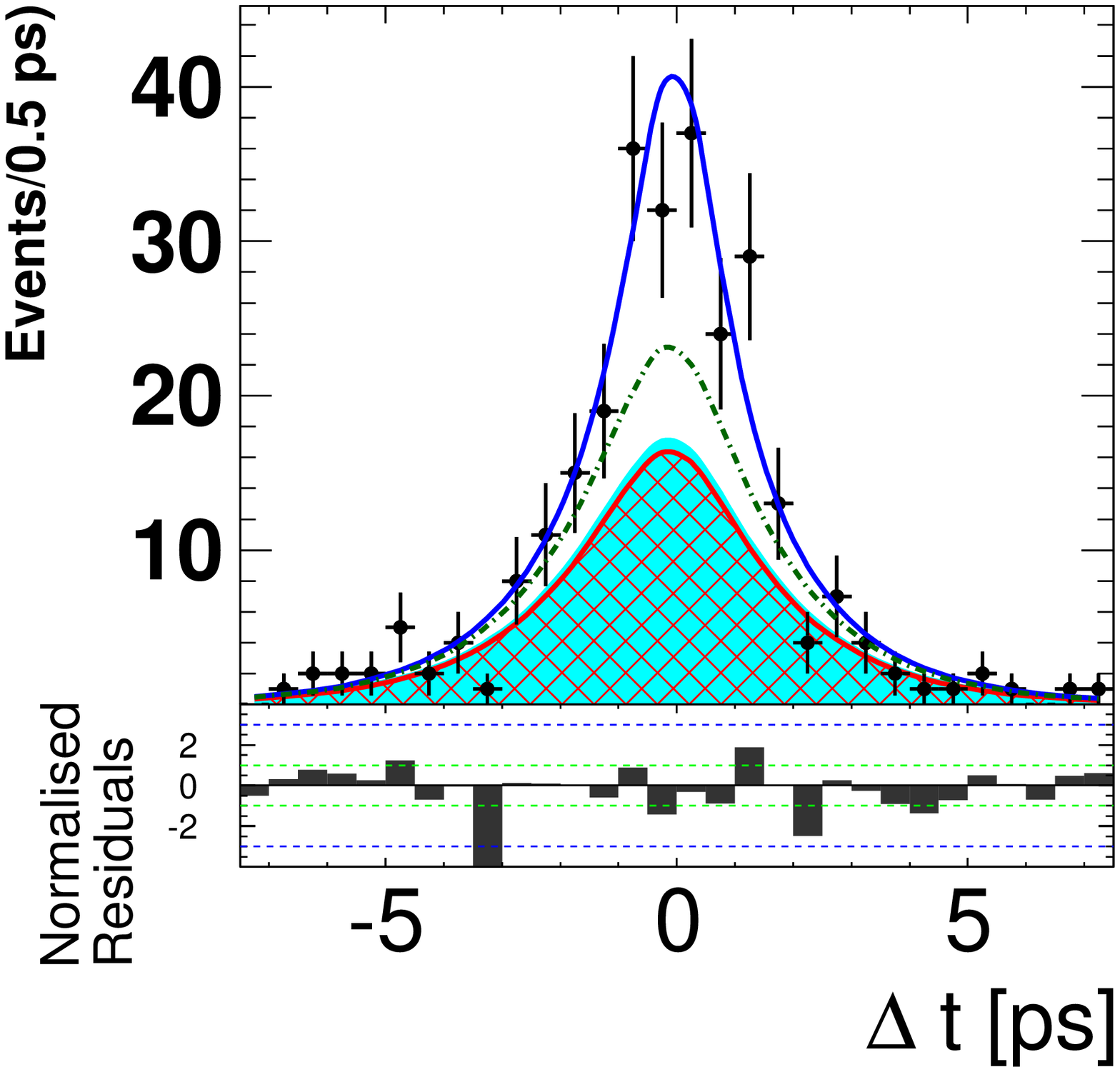}
 \put(-35,120){(h)}
 \caption{(color online) Projections of the fit to the data in
   signal-enhanced regions as described in the text. The (black)
   points represent the data and the solid (blue) curves represent the
   fit result. The hashed (red) areas show the \Brprm\ contribution
   and the bright-shaded (cyan) areas show all four-pion final states.
   (a) and (d): the short-dashed (dark green) curves show the
   non-peaking $B\bar{B}$ contribution and the long-dash (bright
   green) curves show the total non-peaking background. (b), (c), (e)
   and (f): the dash-dotted (dark green) line shows the contribution
   from all $B\bar{B}$ decays. (g): the dash-dotted (magenta) curve
   shows the continuum contribution, the dark (green) area shows the
   entire contribution from $B\bar{B}$ decay. (h): the dashed (dark green) curve shows the non-peaking $B\bar{B}$ contribution. The residuals are plotted below each distribution.}
   \label{fig_bf_data}
\end{figure}

\begin{figure}[h]
  \centering
  \includegraphics[height=300pt,width=!]{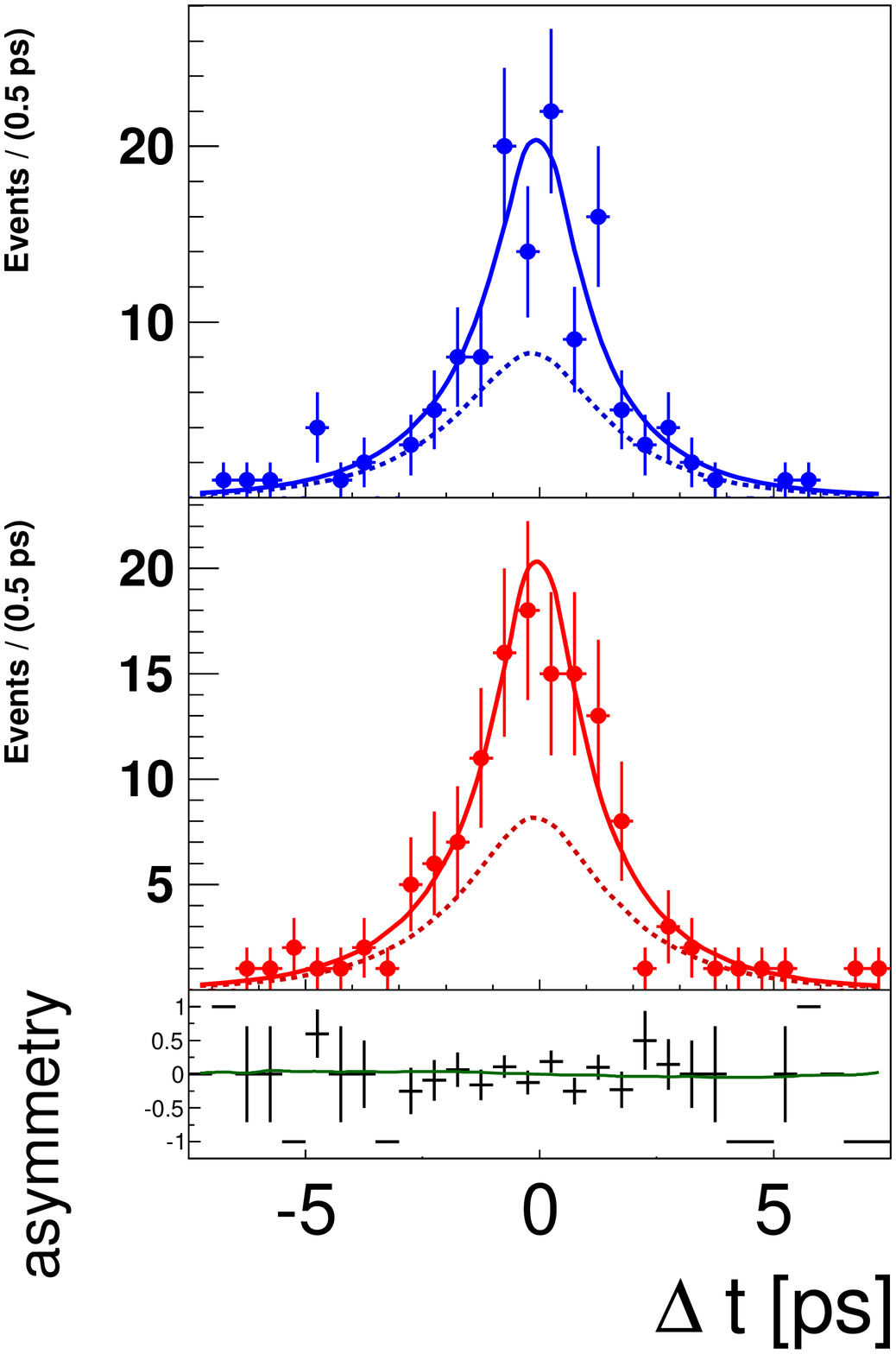}
 \put(-60,270){\textcolor{blue}{q=$+1$}} \put(-60,165){\textcolor{red}{q=$-1$}}\\
 \caption{(color online) Projections of the fit result onto \Dt\ for
   each flavor of \Btag. The points represent the data and the solid curves
   represent the fit result. Signal is shown as a dashed line. The measured asymmetry is plotted below.}
   \label{fig_bf_data_dt}
\end{figure}

From the fit to the data, described in the previous section, we obtain
\begin{eqnarray*}
{\cal B}(\Brprm)&=&(28.3\pm 1.5\;(\rm stat) \pm 1.5\;(\rm syst))\times 10^{-6},\\
f_L &=& 0.988\pm\; 0.012\;(\rm stat ) \pm0.023\;(\rm syst),\\
{\cal A}_{CP} &=&0.00\pm0.10\;(\rm stat ) \pm0.06\;(\rm syst),\\
{\cal S}_{CP} &=&-0.13\pm0.15\;(\rm stat ) \pm0.05\;(\rm syst).
\label{e_fitresultBR}
\end{eqnarray*}
corresponding to $1754\pm94$ and $21\pm 22$ \Brprm\ events with longitudinal and transverse polarization, respectively. The evaluation of the systematical uncertainties given above is described in~\Cref{Systematic Uncertainties}. Signal-enhanced projections of the fit results onto \De, \Mbc, $M_+$, $M_-$, $H_+$, $H_-$ and \Fsb\ are shown in~\Cref{fig_bf_data}, where the signal-enhanced region is defined as $|\De|<0.1 \;{\rm GeV}$, $\Mbc >5.275 \;{\rm GeV}/c^2$, $0.62\;{\rm GeV}/c^2<m(\pi^\pm\piz) <0.92 \;{\rm GeV}/c^2$, $\fevt>0.75$, and $r$-bin$>2$. Depending on the projected variable $10\%$ - $16\%$ of the signal events are retained.\\
A clear signal peak can be seen in the $\De$ and \Mbc\ distributions, while the signal-enhanced projection onto $\Fsb$ remains dominated by the continuum contribution. The $\rho^\pm$ mesons are found to be predominantly in the longitudinally polarized state. Figure~\ref{fig_bf_data_dt} (a) shows the flavor-integrated \Dt\ distribution, including the fit result. No $CP$ violation is observed: the \Dt\ distributions for the two flavors of \Btag\ in~\Cref{fig_bf_data_dt} (b) lacks any asymmetry. The statistical correlation coefficients between the observables are given in~\Cref{tab_corMresult} and are all negligible except for a correlation between ${\cal B}(\Brprm)$ and $f_L$.
The yields of all other four-pion final states are shown in~\Cref{tab_4piY} and are consistent with zero within $90\%$ confidence level except for the $X^0[\pip\pim]\piz\piz$ component, being consistent with combinatorial background. We do not provide the converted branching fractions of the peaking background modes as the systematic uncertainties are expected to be large for the four-pion backgrounds.

\begin{table}[h]                                                                                                                                      
\centering 
 \caption{Correlation matrix as obtained from the fit to the data.}                                                                                                                                       
   \begin{tabular}                                                                                                                                    
     {@{\hspace{0.5cm}}c@{\hspace{0.5cm}}| @{\hspace{0.5cm}}c@{\hspace{0.5cm}} @{\hspace{0.5cm}}c@{\hspace{0.5cm}} @{\hspace{0.5cm}}c@{\hspace{0.5cm}}  @{\hspace{0.5cm}}c@{\hspace{0.5cm}}}                                         \hline \hline                                                                 & ${\cal B}(\Brprm)$ & $f_L$ &${\cal A}_{CP}$ & ${\cal S}_{CP}$\\                                                                   \hline
${\cal B}(\Brprm)$ & 1 & $-$0.228  & $-$0.031 & $-$0.015 \\
$f_L$ & & 1 & 0.003 & 0.026\\
${\cal A}_{CP}$ & & & 1 & 0.018 \\
${\cal S}_{CP}$ & & & & 1 \\
                        
    \hline \hline                                                                                                                                        \end{tabular}                                                                                                                                                                                                                                                                                    
\label{tab_corMresult}                                                                                                                             
\end{table}

\begin{table}
  \centering
  \caption{Yields of the four-pion final states as obtained from the fit to the data. The errors are statistical only. The component $X^0\piz\piz$ is treated as explained in the text, see~\Cref{B4pi model}. The fraction $f_{\rm comb}^{X^0\piz\piz}$ is consistent with one and excludes a significant contribution of $B^0\to X^0[\pip\pim]\piz\piz$ decays. }
  \begin{tabular}
    {@{\hspace{0.5cm}}c@{\hspace{0.5cm}}  @{\hspace{0.5cm}}c@{\hspace{0.5cm}}   @{\hspace{0.5cm}}c@{\hspace{0.5cm}}  } 
    \hline \hline
    Mode & Yield & $f_{\rm comb}^{X^0\piz\piz}$\\
    \hline
  $\Bz \rightarrow a_1^0\piz$ & $86\pm94$ & $--$ \\
 $\Bz \rightarrow (\rho^\pm\pi^\mp\piz)_{\rm NR}$ & $215\pm 131$  & $--$ \\ 
 $\Bz \rightarrow (\pip\piz\pim\piz)_{\rm NR}$ &  $170 \pm 114$ & $--$ \\
$X^0\piz\piz$ & $625 \pm 90$& $1.12\pm0.16$ \\
     \hline \hline
  \end{tabular}
  \label{tab_4piY}
\end{table}

\section{Validity Checks}
\label{Validity Checks and Significance} 

We have validated the fitting procedure by studying a large number of
pseudo experiments, where either all components are generated from the
PDFs or all \BBbar\ components are taken from a full GEANT MC
simulation. Within the statistical error, the fitter reliably recovers
the input values for $f_L$ and ${\cal S}_{CP}$.  For the branching
fraction and ${\cal A}_{CP}$, the fitter exhibits a small bias in the
case of fully simulated events due to imperfections in the modeling of
the variable correlations. The treatment of the bias is described
in~\Cref{Systematic Uncertainties}. The errors from the fit results
are consistent with the expectations from studying fully simulated MC
events. We check the fitting procedure by floating the $B^0$ lifetime
in a separate fit. The obtained lifetime is consistent with the
current world average within one statistical standard deviation.\\
We furthermore perform three checks, where we either require $|r|>0.5$ or $\fevt>0.5$ or where we set the fractions of the signal's 1T and 0T
components to zero in fits to the data. All results are consistent with our baseline result. 
 We perform a fit to data, where we require $0.68\;{\rm GeV}/c^2 < m(\pi^\pm\piz)< 0.86
\;{\rm GeV}/c^2$ to test the assumptions made in the isospin
analysis~\cite{comment_on_phi2}. This removes events with different
$\rho^+$ and $\rho^-$ masses, which breaks the isospin symmetry
assumed in the determination of $\phi_2$. We obtain ${\cal S}_{CP} =-0.09 \pm 0.21\;(\rm stat )$, being consistent
with our nominal result. \\
In order to determine the data-to-simulation correction factors
(see~\Cref{Event Model}), we study a control sample of \drho\ decays,
which are topologically similar to \Brprm. We perform fits including
\De, \Mbc, \fevt, $m(\pi^\pm\piz)$, $\cH^\pm$ and \Dt. The results
obtained for the branching fraction, the fraction of longitudinal
polarized $\rho^\pm$ mesons and also the $B^\pm$ lifetime are in
agreement with the current world averages within one statistical
standard deviation~\cite{PDG}. In addition, the time-dependent
$CP$-violation parameters are consistent with zero when floating them
in the fit to the data.

\section{Systematic Uncertainties}
\label{Systematic Uncertainties} 
Systematic uncertainties from various sources are considered and estimated with independent studies and cross checks and are summarized in~\Cref{t_syst_rprm}. For the total systematic uncertainties, the components are added in quadrature.
They include the uncertainty on the number of produced \BBbar\ events, the track-reconstruction efficiency, the selection efficiency due to particle identification, and the $\piz$ reconstruction, which are determined by using independent control samples. The $\piz$ efficiency correction is obtained from studying $\tau^\pm\to \pi^\pm\piz\nu_\tau$ decays. In the previous Belle measurement, this difference between data and MC was studied with $\eta\to\piz\piz,\;\piz\piz\piz$ decays and was fully treated as a systematic uncertainty.
 Uncertainties affecting
the vertex reconstruction include the interaction point (IP) profile, the track selection based on the track helix
errors, helix parameter corrections, the tag side track selection based on their impact parameters, 
\Dt\ and vertex goodness-of-fit selection, $\Delta z$ bias and SVD
misalignment. 
The \Dt\ resolution function parameters, as well as the flavor tagging performance parameters $w$ and $\Delta w$, are varied within their errors. 
Possible systematic biases from the interference on the tag side arising between the
CKM-favored $b \to c\bar{u}d$ and doubly CKM-suppressed $\bar{b} \to \bar{u}c\bar{d}$ amplitudes in the
final states used for flavor tagging are estimated by studying a large number of MC pseudo-experiments generated with interference~\cite{TSI}. The input parameters for the pseudo-experiments and the amount of their possible changes are estimated by the semileptonic $B$ decay control sample, $B^0 \to D^{*-} l^{+} \nu$.

The parametric model shape is varied within the errors obtained from MC simulation. The fixed ratios listed in~\Cref{tab_bf_fixed} are varied within $\pm 10\%$ and give a negligible contribution. Uncertainties in the non-parametric shapes are obtained by varying the contents of the histogram bins within $\pm1\sigma$ in turn. The fixed physics parameters, the $\rho^\pm$ mass and width as well as the $\tau_{B^0}$ and $\Delta m$, are varied within their world-average uncertainties~\cite{PDG}.
To account for a possible difference in the distributions obtained from MC simulation to data, we vary the fractions of the combinatorial part (reconstruction category $0T$) by $\pm 20\%$ for all components including a $\rho$ resonance in turn.

We account for a difference between data and MC by varying the fraction of the different signal categories by $\pm20\%$ of their values and repeating the fit.
The systematic uncertainty due to fixing the peaking background yield $B\to\omega\piz$ is estimated by varying the branching fraction by its world average error and repeating the fit. For \aonepi\ we vary the yield by a conservative factor of two while also increasing the fraction of correctly reconstructed events. This is motivated by the high sensitivity of the reconstruction efficiency of this decay to the dipion analysis region and a possible difference in the line shape between MC simulation and data.

The fit bias is determined from full simulation by examining the difference between the generated and fitted physics parameters. Because of imperfections in the modeling of the correlations, we find a non-negligible but stable bias of $+0.5\times 10^{-6}$ for the branching fraction of \Brprm\ and of $+0.04$ for ${\cal A}_{CP}$. We study these biases with large statistics ensemble tests, and correct the result. The systematic uncertainties related to the bias correction are obtained from studying pseudo experiments generated with the observables varied within $\pm 1$ statistical standard deviation of the fit result. The largest deviations to the generated values is then taken. All other biases are found to be small compared to the statistical uncertainty and are therefore treated fully as systematic uncertainties.

The dominant uncertainty of the $CP$-violating parameters is due to a possible $CP$ violation in the \BBbar\ backgrounds. We include an asymmetry term in the $\Dt$ PDFs of the \BBbar\ backgrounds and refit the data, where the asymmetry is fixed to $\pm5\%$ $CP$ violation for charm $B$ decays and $\pm50\%$ for charmless $B$ decays in turn. Only direct $CP$ violation is considered for charged $B$ decays and we provide the uncertainties from the four-pion final state backgrounds and from the remaining $B\bar{B}$ decays separately in~\Cref{t_syst_rprm}.

We account for the phase space assumption for $B^{0}\to(\rho^\pm\pi^\mp\piz)_{\rm NR}$ decays by replacing the helicity PDF with one where the two non-resonant pions are either in a S- or a P-wave configuration in the fit to the data. In the latter case, the $\rho^\pm$ can be either longitudinally or transversely polarized. The maximal deviation from the nominal model is taken as the uncertainty related to the assumption of the $B^{0}\to(\rho^\pm\pi^\mp\piz)_{\rm NR}$ helicity dependence. This is the dominant uncertainty in the measurement of $f_L$.

Finally, the uncertainty from neglecting interference between the four-pion final states is estimated by constructing a 4-body amplitude model and studying samples of two four-pion final states, including detector effects. For each set of modes, we first calibrate the relative amplitude strength between two considered modes in order to obtain a yield ratio as found in the data. For the calibration, we set the relative phase to $90^{\circ}$. Then we generate sets where the relative phase between the two modes of interest varies from $0^{\circ}$ to $180^{\circ}$ in steps of $10^{\circ}$. Each set is fitted with an incoherent model and the largest root-mean-square error of the variation of the fit results is taken to be the systematic uncertainty for each observable. We consider the modes $B^{0}\to \rop\rom, a_1^\pm\pi^\mp, (\rho^\pm\pi^\mp\piz)_{\rm NR}$, and $(\pip\piz\pim\piz)_{\rm NR}$, and find that interference is almost negligible.

Due to the large variety of backgrounds, changes in the model tend to affect the background yields, while leaving the signal yield rather stable.

 \begin{table}[h]                                                                                                                                                               
\centering
\caption{Systematic uncertainties.}
                                                                                                                                                          
 \begin{tabular}{@{\hspace{0.5cm}}c@{\hspace{0.5cm}} | @{\hspace{0.5cm}}c@{\hspace{0.5cm}} @{\hspace{0.5cm}}c@{\hspace{0.5cm}} @{\hspace{0.5cm}}c@{\hspace{0.5cm}} @{\hspace{0.5cm}}c@{\hspace{0.5cm}}}
\hline \hline 
\small Category & $\delta{\cal B}$($\%$) & $\delta f_L$ & $\delta {\cal A}_{CP}[10^{-2}]$ & $\delta {\cal S}_{CP}[10^{-2}]$\\[1ex]
\hline \hline 
$N(B\bar{B})$ & 1.38 & - & -& -\\[1ex]
Tracking & 0.70 &  -& -&-\\[1ex]
PID & 2.50 &  -& -&-\\[1ex]
$\piz$ reconstruction & 2.98 &  -& -&-\\[1ex]     
\hline
IP profile & 0.01 & 0.001 & 0.68 & 0.94\\[1ex] 
$\Dt$ selection  & 0.00 & 0.001 & 0.04 & 0.06\\[1ex]  
Track helix error  & 0.00 & 0.000 & 0.02 & 0.01 \\[1ex]  
Vertex quality & 0.16 & 0.000 & 1.20 & 0.60 \\[1ex] 
Tagside track selection & 0.01 & 0.001 & 0.84 & 0.95 \\[1ex] 
$\Delta z$ bias &-  &- & 0.50& 0.40 \\[1ex] 
Misalignement & - &- &0.40 &0.20 \\[1ex] 
Resolution function & 0.00 &0.000 &0.00 &0.00 \\[1ex] 
Flavor tagging & 0.07 & 0.002 & 0.71& 0.51\\[1ex] 
Tagside interference & - &- & 1.02& 0.08 \\[1ex] 
\hline 
Model shape & 3.47 &0.003 & 0.30 &0.60 \\[1ex]
Histogram shape & 0.17 & 0.002 & 0.19 & 0.31\\[1ex]
Physics parameters & 0.00 &0.000 & 0.02 & 0.02\\[1ex]
MC composition & 0.04  & 0.007 & 0.64 & 1.34 \\[1ex]  
Misreconstructed fraction & 0.01 & 0.001 & 0.60 & 0.50 \\[1ex]
Fixed background yields & 0.00 & 0.001 &0.04 &0.08 \\[1ex]   
$B\to a_1^\pm\pi^\mp$ description & 0.01& 0.002& 0.09&0.20 \\[1ex]
Fit bias & 0.53 & 0.002 & 0.50 & 0.74\\[1ex]
Background $CP$ violation & 0.00 & 0.000 & 4.92 & 2.75\\[1ex]    
$B\to 4\pi$ $CP$ violation & 0.03 & 0.006 & 3.03 & 3.65\\[1ex]    
$(\rho^\pm\pi^\mp\piz)_{\rm NR}$ helicity & 0.04 &0.020 & 0.12 &0.77 \\[1ex]
Interference & 0.01 & 0.002 & 0.12 & 0.15 \\[1ex]
 \hline \hline
Total & 5.47 &0.023 & 6.37 & 5.42 \\[1ex]
(absolute uncertainty $[\times 10^{6}]$) & 1.55 & & & \\[1ex]
\hline \hline  
 \end{tabular}                                          
\label{t_syst_rprm}                                                                                                                                                           
\end{table}     

\clearpage

\section{Constraints on the CKM angle ${\boldmath\phi_{2}}$}
\label{phi2 constraint}

As stated in the Introduction, the $CP$ violating parameters of the decay \Brprm\ obtained in this paper can be used to constrain the angle $\phi_2$ in the CKM unitarity triangle. In the following, we estimate the possible pollution from loop diagrams with two methods: isospin invariance and SU(3) flavor symmetry.

\subsection{Isospin}

\begin{figure}[h]                                                                                                                                   
\centering                                                                                                                                           
\includegraphics[height=!,width=0.45\columnwidth]{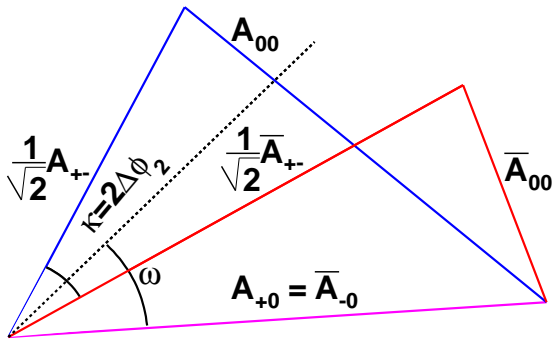}
\caption{Sketch of the isospin triangles for $B$ and $\bar{B}$ decays into unflavored final states with isospin I=1.}
\label{p_phi2_tr}                                                                                                                                                   
\end{figure}

\begin{figure}[h]                                                                                                                                   
\centering                                                                                                                                           
\includegraphics[height=!,width=0.45\columnwidth]{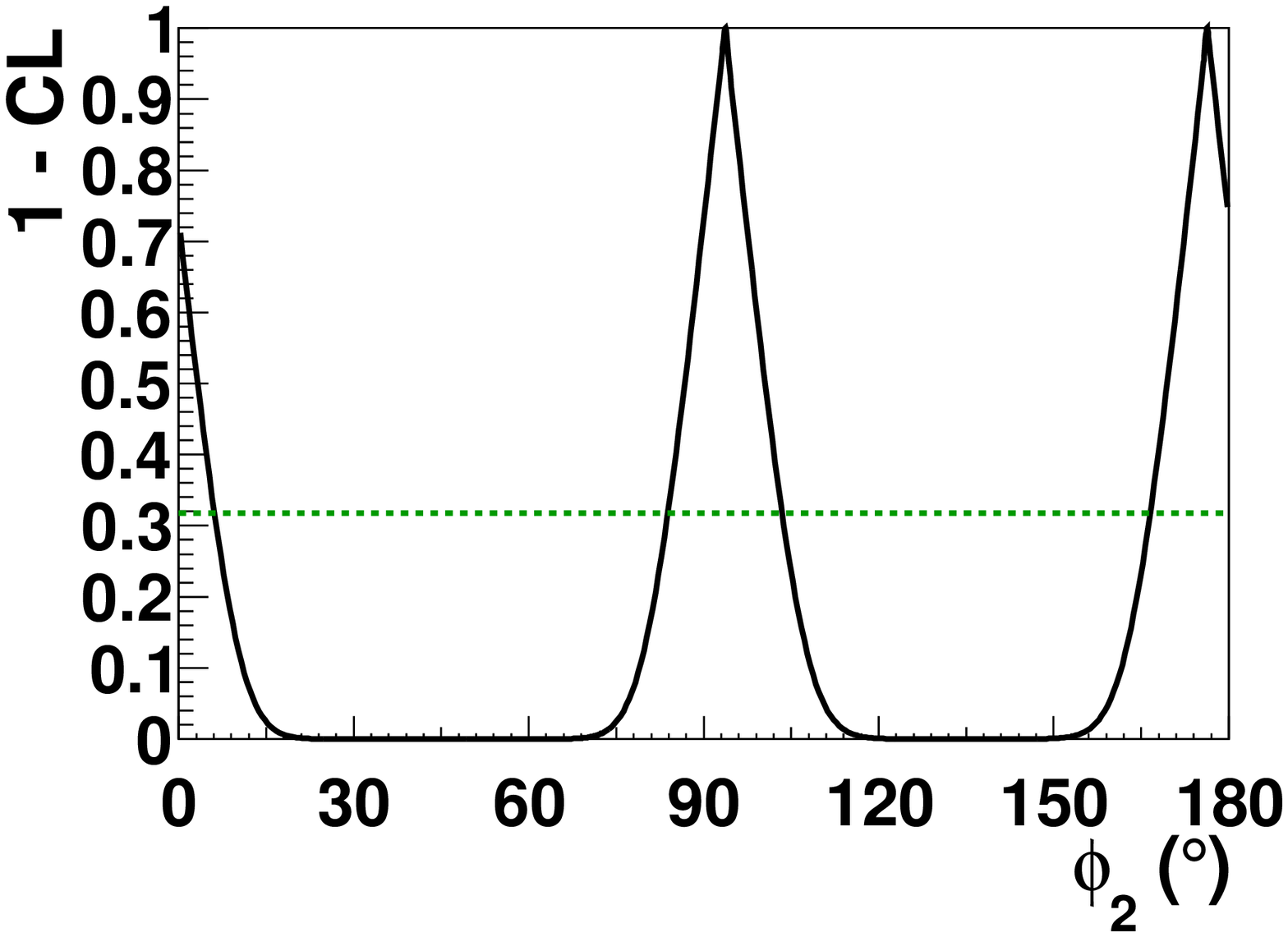}
\includegraphics[height=!,width=0.45\columnwidth]{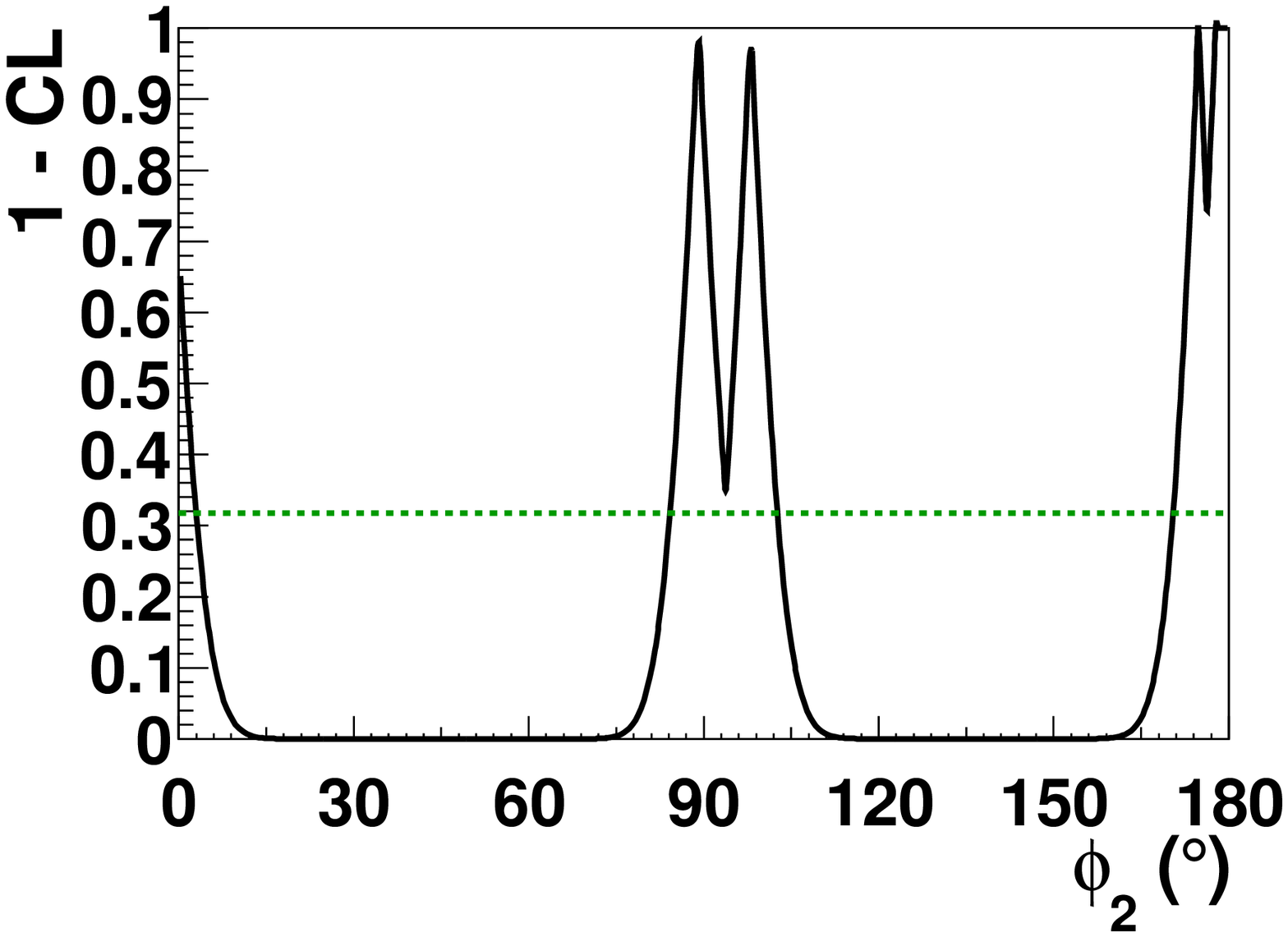}
\put(-380,120){(a)}\put(-165,120){(b)}
\caption{$1-$CL versus $\phi_2$ obtained from $B\to\rho\rho$ decays. (a) isospin analysis, (b) SU(3) flavor analysis. The horizontal line shows the $68\%$ CL.}
\label{p_phi2}                                                    
\end{figure}

We use our result together with other Belle results to obtain a constraint on $\phi_2$ from a isospin analysis~\cite{theory_isospin} in the $B\to\rho\rho$ system. Neglecting electroweak contributions or isospin-breaking effects, the complex $B\to\rho\rho$ amplitudes for the various charge configurations of the $\rho\rho$ system can be related via
\begin{equation}
\frac{1}{\sqrt{2}}A_{+-} + A_{00} = A_{+0}, \;\;\;\;\;\;\;\;\frac{1}{\sqrt{2}}\bar{A}_{+-} + \bar{A}_{00} = \bar{A}_{-0},
\end{equation}
where the amplitudes with $\bar{b}\to \bar{u}$ ($b \to u$) transitions
are denoted as $A_{ij}$ ($\bar{A}_{ij}$) and the subscripts identify
the charges of the two $\rho$ mesons. These relations can be
visualized as two isospin triangles, as shown in~\Cref{p_phi2_tr}.
 Since the charged $B$ decay $B^{\pm}\to\rho^{\pm}\rho^0$ proceeds
 only via a tree level diagram, the two isospin triangles share the
 same base $A_{+0} = \bar{A}_{-0}$. The phase difference between the
 two sides $A_{+-}$ and $\bar{A}_{+-}$ corresponds to the shift $2
 \Delta \phi_2$ due to additional contributions. This method leads to
 an eight-fold ambiguity in the determination of $\phi_2$, as there
 are four possible orientations of the two triangles and two solutions
 from the trigonometric conversion of $\sin(\phi_2^{\rm eff})$. The amplitudes are constructed from the branching fractions and the direct $CP$ asymmetries ${\cal A}_{CP}$ and are then used to obtain the possible pollution in the mixing-induced $CP$ asymmetry ${\cal S}_{CP} = \sqrt{1-A_{CP}^{2}} \sin(2\phi_{2}^{\rm eff})$, obtained from the measurement of $B^0\to\rho^+\rho^-$ decays. The remaining sides of the triangles are constructed from other Belle results: the longitudinally polarized fraction of ${\cal B}(B^{0}\to \rho^{0}\rho^{0}) = (1.02 \pm 0.34) \times 10^{-6}$ with $f_L^{00} = 0.21\pm 0.25$~\cite{Br0r0_pit} and the longitudinally polarized fraction of ${\cal B}(B^{\pm}\to \rho^{\pm}\rho^{0}) = (31.7 \pm 8.8)\times 10^{-5}$ with $f_L^{+0} = 0.95\pm0.11$~\cite{rpr0_Belle}. We convert the $\chi^2$ distribution constructed from the five amplitudes, including the correlations obtained from this measurement, into a probability scan as shown in~\Cref{p_phi2} (a). Two solutions for $\phi_2$ are found, the one that is consistent with other SM-based constraints yields $\phi_{2} = (93.7 \pm 10.6)^{\circ}$. The size of the penguin contributions is consistent with zero: $\Delta \phi_2 = (0.0 \pm 9.6)^{\circ}$. Because of the very small $\Brr$ branching fraction relative to the other two $B\to\rho\rho$ decays, the four solutions from the isospin analysis collapse into the two distinct solutions.

\subsection{SU(3) Flavor}
The amplitude of \Brprm\ decays can be written in terms of tree and penguin contributions
\begin{equation}
{\cal A}_{\Brprm} = Te^{i\phi_3} + Pe^{i\delta_{PT}},
\end{equation} 
where $T$ and $P$ are the magnitude of the tree and penguin amplitudes, respectively, $\delta_{PT}$ is their strong-phase difference and $\phi_3$ is the phase of $V_{ub}$.
Following Ref.~\cite{theory_SU3}, the SU(3) symmetry provides an alternative way to remove the penguin contribution from $\phi_{2}^{\rm eff}$ by relating \Brprm\ decays to the pure penguin mode $B^+ \to K^{0*}\rho^+$:
\begin{align}
\frac{{\cal B}_{\rm LP}(B^+\to K^{*0}\rop)}{{\cal B}_{\rm LP}(\Brprm)}
= \frac{\tau_{B^{\pm}}}{\tau_{B^{0}}}\biggl(\frac{|V_{cs}|f_{K^*}}{|V_{cd}|f_{\rho}}\biggr)^{2} \frac{Fr_{PT}^2}{1-2r_{PT}\cos\delta_{PT}\cos(\phi_1+\phi_2) + r_{PT}^2},
\end{align}
where $r_{PT}=|P|/|T|$ and the factor $F=0.9\pm0.6$ is taken from Ref.~\cite{theory_SU3} and accounts for possible SU(3) breaking ($F=1$ corresponds to no breaking). ${\cal B}_{\rm LP}$ denotes the branching fractions for longitudinal polarization; $\tau_{B^\pm}$ and  $\tau_{B^0}$ the $B^\pm$ and $B^0$ lifetimes, respectively; $V_{ij}$ is a CKM matrix element; and $f_{k}$ is a form factor. The CKM phase $\phi_1$ is taken from the measurement of $b\to c\bar{c}s$ decays~\cite{jpsiks_Belle2} and $B^+ \to K^{0*}\rho^+$ related inputs from~\cite{Belle_Kstarrhop}.
With
\begin{align}
-{\cal A}_{CP} &= \frac{2r_{PT}\sin\delta_{PT}\sin(\phi_1 +
  \phi_2)}{1-2r_{PT}\cos\delta_{PT}\cos(\phi_1+\phi_2) + r_{PT}^2}\\
{\cal S}_{CP} &= \frac{\sin 2\phi_2 +
  2r_{PT}\cos\delta_{PT}\sin(\phi_1 - \phi_2) - r_{PT}^2\sin2\phi_1}{1-2r_{PT}\cos\delta_{PT}\cos(\phi_1+\phi_2) + r_{PT}^2},
\end{align}
a probability scan similar to the isospin analysis can be performed, as shown in~\Cref{p_phi2} (b), where each of the two double peaks consists of one solution for $\delta_{PT}<90^\circ$ and one otherwise. The solution most compatible with other SM-based constraints and for the theoretically motivated case  $\delta_{PT}<90^\circ$ is $\phi_2=(89.3 \pm 4.8\;(\rm scan)^{+1.0}_{-3.4}\;(\rm SU(3)))^\circ$. Varying $F$ within the quoted error results in the second uncertainty, denoted as ``SU(3)''. We furthermore obtain $r_{PT} = 0.09 \pm 0.02\; (\rm scan) ^{+0.06}_{-0.02}\; (\rm SU(3))$ and $\delta_{PT} = (0.0 \pm 48.7 \;(\rm scan) \pm 0.0\;(\rm SU(3)))^{\circ}$.

\section{Conclusions}
\label{Conclusion}
We have presented measurements of the branching fraction of \Brprm\ decays, the fraction of longitudinally polarized $\rho^\pm$ mesons, as well as the $CP$-violating parameters in the decay into a pair of longitudinally polarized $\rho$ mesons using the final Belle data set of $772\times 10^{6}$ \BBbar\ pairs. Improvements compared to previous Belle measurements are the increased data sample and the simultaneous extraction of all observables. The measurement is optimized for a high signal yield, as the $CP$-violating parameters are still statistically limited. The inclusion of the helicity angles provides additional and strong separation power between the various components and the simultaneous fit to $\Dt$ improves the continuum separation in particular. This procedure reduces the statistical uncertainties significantly at the cost of higher analysis complexity and a longer computing time. The obtained results are in excellent agreement with the previous measurements~\cite{Brprm_Belle,rhorho_Belle,rhorho_BABAR} and predictions~\cite{Bm1m2_NNLO,fQCD} and are currently the most precise measurement of the branching fraction and longitudinal polarization fraction as well as the tightest constraint on $CP$ violation in this decay. We use our results, together with other Belle measurements of $B\to\rho\rho$ decays, to constrain the internal angle $\phi_2$ of the CKM unitarity triangle with an isospin analysis. We obtain $\phi_{2} = (93.7 \pm 10.6)^{\circ}$ as the solution most compatible with other SM-based fits. The uncertainty of this scan is dominated by the large uncertainty on the branching fraction for the decay $B^\pm\to\rho^\pm\rho^0$. We provide an alternative constraint on $\phi_2$ by exploiting the SU(3) flavor symmetry and obtain $\phi_2=(89.3 \pm 4.8\;(\rm scan)^{+1.0}_{-3.4}\;(\rm SU(3)))^\circ$ for the theoretically motivated case of $\delta_{PT}<90^{\circ}$.

\section*{Acknowledgments}

We thank the KEKB group for the excellent operation of the
accelerator; the KEK cryogenics group for the efficient
operation of the solenoid; and the KEK computer group,
the National Institute of Informatics, and the 
PNNL/EMSL computing group for valuable computing
and SINET4 network support.  We acknowledge support from
the Ministry of Education, Culture, Sports, Science, and
Technology (MEXT) of Japan, the Japan Society for the 
Promotion of Science (JSPS), and the Tau-Lepton Physics 
Research Center of Nagoya University; 
the Australian Research Council and the Australian 
Department of Industry, Innovation, Science and Research;
Austrian Science Fund under Grant No.~P 22742-N16 and P 26794-N20;
the National Natural Science Foundation of China under Contracts 
No.~10575109, No.~10775142, No.~10875115, No.~11175187, and  No.~11475187;
the Chinese Academy of Science Center for Excellence in Particle Physics; 
the Ministry of Education, Youth and Sports of the Czech
Republic under Contract No.~LG14034;
the Carl Zeiss Foundation, the Deutsche Forschungsgemeinschaft
and the VolkswagenStiftung;
the Department of Science and Technology of India; 
the Istituto Nazionale di Fisica Nucleare of Italy; 
National Research Foundation (NRF) of Korea Grants
No.~2011-0029457, No.~2012-0008143, No.~2012R1A1A2008330, 
No.~2013R1A1A3007772, No.~2014R1A2A2A01005286, No.~2014R1A2A2A01002734, 
No.~2014R1A1A2006456;
the Basic Research Lab program under NRF Grant No.~KRF-2011-0020333, 
No.~KRF-2011-0021196, Center for Korean J-PARC Users, No.~NRF-2013K1A3A7A06056592; 
the Brain Korea 21-Plus program and the Global Science Experimental Data 
Hub Center of the Korea Institute of Science and Technology Information;
the Polish Ministry of Science and Higher Education and 
the National Science Center;
the Ministry of Education and Science of the Russian Federation and
the Russian Foundation for Basic Research;
the Slovenian Research Agency;
the Basque Foundation for Science (IKERBASQUE) and 
the Euskal Herriko Unibertsitatea (UPV/EHU) under program UFI 11/55 (Spain);
the Swiss National Science Foundation; the Ministry of Education and the Ministry of Science and
 Technology of Taiwan; and the U.S.\
Department of Energy and the National Science Foundation.
This work is supported by a Grant-in-Aid from MEXT for 
Science Research in a Priority Area (``New Development of 
Flavor Physics'') and from JSPS for Creative Scientific 
Research (``Evolution of Tau-lepton Physics'').

\section*{Appendix}
\label{appendix}
\subsection{Signal Correlation Matrices}

   \begin{table}[tbh]
\centering
\begin{tabular}
{@{\hspace{0.25cm}}c@{\hspace{0.25cm}} | @{\hspace{0.25cm}}c@{\hspace{0.25cm}} @{\hspace{0.25cm}}c@{\hspace{0.25cm}} @{\hspace{0.25cm}}c@{\hspace{0.25cm}} @{\hspace{0.25cm}}c@{\hspace{0.25cm}} @{\hspace{0.25cm}}c@{\hspace{0.25cm}} @{\hspace{0.25cm}}c@{\hspace{0.25cm}} @{\hspace{0.25cm}}c@{\hspace{0.25cm}} @{\hspace{0.25cm}}c@{\hspace{0.25cm}} }
\hline \hline
0 & \De\ & $M_{bc}$  & $\Mpp^{1}$ & $\Mpp^{2}$ &  \fevt  & $\cH^{1}$ & $\cH^{2}$ & $\Delta t$\\ 
\hline
\De  &  1 & -0.01 & 0.04 & 0.04 & 0.00 &  -0.09 & -0.09 & 0.00 \\
$M_{bc}$ &  &  1 & -0.00 & -0.00 & 0.00 &  -0.02 & -0.02 & 0.00 \\
$\Mpp^{1}$ & &  &  1  & -0.00 & -0.03 & -0.02 & -0.00 & 0.00 \\
$\Mpp^{2}$ &  &  &    & 1     & -0.02 & 0.00 & -0.02 & -0.00 \\
\fevt   &   &   &    &       & 1      & 0.00 & 0.01 & -0.00 \\
$\cH^{1}$  &   &  &     &       &        & 1      & -0.04 & -0.00  \\
$\cH^{2}$  &   &   &    &       &        &        & 1 & 0.00 \\
$\Delta t$  &   &   &    &       &        &        &  &  1  \\
\hline \hline
\end{tabular}
\caption{Correlation matrix for the truth model (LP, SVD2).}
\label{t_corsigLP2}
\end{table}

   \begin{table}[tbh]
\centering
\begin{tabular}
{@{\hspace{0.25cm}}c@{\hspace{0.25cm}} | @{\hspace{0.25cm}}c@{\hspace{0.25cm}} @{\hspace{0.25cm}}c@{\hspace{0.25cm}} @{\hspace{0.25cm}}c@{\hspace{0.25cm}} @{\hspace{0.25cm}}c@{\hspace{0.25cm}} @{\hspace{0.25cm}}c@{\hspace{0.25cm}} @{\hspace{0.25cm}}c@{\hspace{0.25cm}} @{\hspace{0.25cm}}c@{\hspace{0.25cm}} @{\hspace{0.25cm}}c@{\hspace{0.25cm}} }
\hline \hline
0 & \De\ & $M_{bc}$  & $\Mpp^{1}$ & $\Mpp^{2}$ &  \fevt  & $\cH^{1}$ & $\cH^{2}$ & $\Delta t$\\ 
\hline
\De  &  1 & 0.00 & 0.06 & 0.06 & 0.05 &  -0.03 & -0.04 & 0.00 \\
$M_{bc}$ &  &  1 & -0.03 & -0.02 & 0.03 &  0.03 & 0.02 & 0.00 \\
$\Mpp^{1}$ & &  &  1  & -0.02 & -0.01 & -0.11 & 0.08 & 0.00 \\
$\Mpp^{2}$ &  &  &    & 1     & -0.02 & 0.08 & -0.12 & -0.01 \\
\fevt   &   &   &    &       & 1      & 0.02 & 0.02 & 0.00 \\
$\cH^{1}$  &   &  &     &       &        & 1      & -0.40 & 0.00  \\
$\cH^{2}$  &   &   &    &       &        &        & 1 & -0.00 \\
$\Delta t$  &   &   &    &       &        &        &  &  1  \\
\hline \hline
\end{tabular}
\caption{Correlation matrix for the 2T signal model (LP, SVD2).}
\label{t_cor_nopi0LP_svd2}
\end{table}

   \begin{table}[tbh]
\centering
\begin{tabular}
{@{\hspace{0.25cm}}c@{\hspace{0.25cm}} | @{\hspace{0.25cm}}c@{\hspace{0.25cm}} @{\hspace{0.25cm}}c@{\hspace{0.25cm}} @{\hspace{0.25cm}}c@{\hspace{0.25cm}} @{\hspace{0.25cm}}c@{\hspace{0.25cm}} @{\hspace{0.25cm}}c@{\hspace{0.25cm}} @{\hspace{0.25cm}}c@{\hspace{0.25cm}} @{\hspace{0.25cm}}c@{\hspace{0.25cm}} @{\hspace{0.25cm}}c@{\hspace{0.25cm}} }
\hline \hline
0 & \De\ & $M_{bc}$  & $\Mpp^{1}$ & $\Mpp^{2}$ &  \fevt  & $\cH^{1}$ & $\cH^{2}$ & $\Delta t$\\ 
\hline
\De  &  1 & 0.01 & 0.01 & 0.02 & 0.04 &  -0.09 & -0.09 & -0.00 \\
$M_{bc}$ &  &  1 & -0.02 & -0.02 & 0.03 &  0.02 & 0.03 & -0.01 \\
$\Mpp^{1}$ & &  &  1  & -0.01 & -0.00 & 0.04 & -0.08 & 0.00 \\
$\Mpp^{2}$ &  &  &    & 1     & -0.01 & -0.08 & 0.06 & 0.00 \\
\fevt   &   &   &    &       & 1      & 0.04 & 0.03 & 0.01 \\
$\cH^{1}$  &   &  &     &       &        & 1      & -0.39 & 0.01  \\
$\cH^{2}$  &   &   &    &       &        &        & 1 & 0.01 \\
$\Delta t$  &   &   &    &       &        &        &  &  1  \\
\hline \hline
\end{tabular}
\caption{Correlation matrix for the signal 1T model (LP, SVD2).}
\label{t_cor1cpip_svd2}
\end{table}

   \begin{table}[tbh]
\centering
\begin{tabular}
{@{\hspace{0.25cm}}c@{\hspace{0.25cm}} | @{\hspace{0.25cm}}c@{\hspace{0.25cm}} @{\hspace{0.25cm}}c@{\hspace{0.25cm}} @{\hspace{0.25cm}}c@{\hspace{0.25cm}} @{\hspace{0.25cm}}c@{\hspace{0.25cm}} @{\hspace{0.25cm}}c@{\hspace{0.25cm}} @{\hspace{0.25cm}}c@{\hspace{0.25cm}} @{\hspace{0.25cm}}c@{\hspace{0.25cm}} @{\hspace{0.25cm}}c@{\hspace{0.25cm}} }
\hline \hline
0 & \De\ & $M_{bc}$  & $\Mpp^{1}$ & $\Mpp^{2}$ &  \fevt  & $\cH^{1}$ & $\cH^{2}$ & $\Delta t$\\ 
\hline
\De  &  1 & -0.01 & 0.00 & 0.03 & 0.06 &  -0.05 & -0.04 & -0.02 \\
$M_{bc}$ &  &  1 & 0.00 & -0.02 & 0.03 &  0.05 & 0.05 & 0.01 \\
$\Mpp^{1}$ & &  &  1  & 0.05 & 0.01 & 0.01 & -0.00 & -0.01 \\
$\Mpp^{2}$ &  &  &    & 1     & -0.00 & 0.01 & -0.01 & 0.01 \\
\fevt   &   &   &    &       & 1      & 0.09 & 0.10 & -0.00 \\
$\cH^{1}$  &   &  &     &       &        & 1      & -0.01 & 0.00  \\
$\cH^{2}$  &   &   &    &       &        &        & 1 & -0.01 \\
$\Delta t$  &   &   &    &       &        &        &  &  1  \\
\hline \hline
\end{tabular}
\caption{Correlation matrix for the signal 0T model (LP, SVD2).}
\label{t_corSC_svd2}
\end{table}


\newpage
\begin{thebibliography}{99}

\bibitem{Cabibbo}
  N.~Cabibbo, Phys. Rev. Lett. {\bf 10}, 531 (1963).

\bibitem{KM}
  M.~Kobayashi and T.~Maskawa, Prog. Theor. Phys. {\bf 49}, 652 (1973).


\bibitem{jpsiks_Belle}
  K.~Abe {\it et al.} (Belle Collaboration), Phys. Rev. Lett. {\bf 87}, 091802 (2001).
\bibitem{jpsiks_Belle2}
I.~Adachi {\it et al.} (Belle Collaboration), Phys. Rev. Lett. {\bf 108}, 171802 (2012). 
\bibitem{jpsiks_BABAR}
  B.~Aubert {\it et al.} (BaBar Collaboration), Phys. Rev. Lett. {\bf 87}, 091801 (2001). 
\bibitem{jpsiks_BABAR2}
  B.~Aubert {\it et al.} (BaBar Collaboration), Phys. Rev. D {\bf 79}, 072009 (2009).

\bibitem{CCimplied}
Here and in the following the charge-conjugated transition is implied unless otherwise stated.

\bibitem{BelleResults}
J.~Brodzicka {\it et al.}, Prog.~Theor.~Exp.~Phys. {\bf 2012}, 04D001 (2012).

\bibitem{CKMandUT}
M.~Battaglia, A.~J.~Buras, P.~Gambino, A.~Stocchi, {\it et al.}, arXiv:hep-ph/0304132.

\bibitem{ckmfitter}
J. Charles {\it et al.} (CKMfitter Group), Eur. Phys. J. C41, 1-131 (2005) [hep-ph/0406184], updated results and plots available at: http://ckmfitter.in2p3.fr .


\bibitem{pipi_Belle}
  H.~Ishino {\it et al.} (Belle Collaboration), Phys.~Rev.~Lett. {\bf 98}, 211801 (2007).

\bibitem{pipi_BABAR}
  B.~Aubert {\it et al.} (BaBar Collaboration),  Phys. Rev. D {\bf 87}, 052009 (2013).

\bibitem{pipi_LHCb}
  R.~Aaij  {\it et al.} (LHCb Collaboration),  JHEP {\bf 1210}, 037 (2012). 




\bibitem{rhopi_Belle}
  A.~Kusaka {\it et al.} (Belle Collaboration), Phys.~Rev.~Lett. {\bf 98}, 221602 (2007).

\bibitem{rhopi_BABAR}
  B.~Aubert {\it et al.} (BaBar Collaboration), Phys.~Rev.~D {\bf 76}, 012004 (2007).



\bibitem{Brprm_Belle}
  A.~Somov {\it et al.} (Belle Collaboration), Phys.~Rev.~Lett. {\bf 96}, 171801 (2006).
\bibitem{rhorho_Belle}
  A.~Somov {\it et al.} (Belle Collaboration), Phys.~Rev.~D {\bf 76}, 011104 (2007).

\bibitem{rhorho_BABAR}
  B.~Aubert {\it et al.} (BaBar Collaboration), Phys.~Rev.~D {\bf 76}, 052007 (2007).

\bibitem{Br0r0_pit}
P.Vanhoefer {\it et al.} (Belle Collaboration), Phys. Rev. D {\bf 89}, 072008 (2014).
\bibitem{r0r0_BABAR}
  B.~Aubert {\it et al.} (BaBar Collaboration), Phys.~Rev.~D {\bf 78}, 071104 (2008).

\bibitem{jeremy_a1pi}
J.~Dalseno  {\it et al.} (Belle Collaboration), Phys.~Rev.~D {\bf 86}, 092012 (2012).

\bibitem{a1pi_BABAR1}
  B.~Aubert {\it et al.} (BaBar Collaboration), Phys.~Rev.~Lett. {\bf 97}, 051802 (2006).

\bibitem{a1pi_BABAR2}
  B.~Aubert {\it et al.} (BaBar Collaboration), Phys.~Rev.~Lett. {\bf 98}, 181803 (2007).

\bibitem{theory_isospin}
  M.~Gronau and D.~London, Phys.~Rev.~Lett. {\bf 65}, 3381 (1990).

\bibitem{theory_SU3}
M. Beneke, M.Gronau, S. Jaeger and M. Spranger, Phys. Lett. B {\bf 638}, 68 (2006).


\bibitem{pQCD}
H.~N.~Li and S.~Mishima, Phys. Rev. D {\bf 73}, 114014 (2006).

\bibitem{fQCD2}
W.~Zou and Z.~Xiao, Phys. Rev. D {\bf 72}, 094026 (2005).

\bibitem{fQCD3}
M.~Beneke, G.~Buchalla, M.~Neubert, C.~T.~Sachrajda, Nucl.~Phys.~B {\bf 591}, 313 (2000).

\bibitem{fQCD4}
 G.~Buchalla, Heavy Quark Theory, CERN-TH/2002-018, arXiv:hep-ph/0202092.

\bibitem{BVV}
M.~Bartsch, G.~Buchalla, C.~Kraus, arXiv:0810.0249v1.

\bibitem{fQCD}
M.~Beneke, J.~Rohrer and D.~Yang, Nucl.~Phys.~B {\bf 774}, 64 (2007).

\bibitem{BVV2}
H.Y.~Cheng, K.C.~Yang, Phys.~Rev.~D {\bf 78}, 094001 (2008).

\bibitem{Bm1m2_NNLO}
G.Bell, V.Pilipp, Phys. Rev. D {\bf 80}, 054024 (2009).

\bibitem{KEKB}
  S.~Kurokawa and E.~Kikutani, Nucl. Instr. and Meth. Phys. Res. Sect.
 A {\bf 499}, 1 (2003), and other papers included in this Volume;
 T.Abe {\it et al.}, Prog. Theor. Exp. Phys. {\bf 2013}, 03A001 (2013) and following articles up to 03A011. 

\bibitem{Belle}
A.~Abashian {\it et al.} (Belle Collaboration), Nucl. Instr. and Meth. 
 Phys. Res. Sect. A {\bf 479}, 117 (2002), also see detector section in
 J.Brodzicka {\it et al.}, Prog. Theor. Exp. Phys. {\bf 2012}, 04D001 (2012).

\bibitem{svd2} Z.~Natkaniec {\it et al.} (Belle SVD2 Group), Nucl. Instr. and Meth.. A {\bf 560}, 1 (2006).

\bibitem{GEANT}
  R.~Brun {\it et al.}, GEANT 3.21, CERN DD/EE/84-1 (1984).


\bibitem{ResFunc}
  H. Tajima {\it et al.}, Nucl. Instr. and Meth. A {\bf 533}, 370 (2004). 




\bibitem{PDG}
 K.A. Olive {\it et al.} (Particle Data Group), Chin. Phys. C {\bf 38}, 090001 (2014). 

\bibitem{fisher}
R.~A.~Fisher, Annals of Human Genetics {\bf 7}, 179 (1936).

\bibitem{foxw1} 
G.C.~Fox and S.~Wolfram, Phys. Rev. Lett. {\bf 41}, 1581 (1978).
 \bibitem{foxw2}
K.~Abe {\it et al.} (Belle Collaboration), Phys. Lett. B {\bf 511}, 151 (2001).


\bibitem{Tagging}
  H.~Kakuno {\it et al.} (Belle Collaboration), Nucl. Instrum. and Meth. A {\bf 533}, 516 (2004). 

\bibitem{pit_phd}
P.~Vanhoefer, Study of $B^0 \to \rho\rho$ decays with the Belle experiment.  Dissertation, LMU M\"unchen: Faculty of Physics, urn:nbn:de:bvb:19-183537 (2015).

\bibitem{BWFF}
J. M. Blatt and V. F. Weisskopf, Theoretical Nuclear Physics (Wiley, New York, 1952).

\bibitem{argus}
H. Albrecht {\it et al.} (ARGUS Collaboration), Phys. Lett. B {\bf 241}, 278 (1990).

\bibitem{TSI}
O.~Long, M.~Baak, R.N.~Cahn, D.~Kirkby, Phys. Rev. D {\bf 68}, 034010 (2003).

\bibitem{comment_on_phi2}
A. F. Falk, Z. Ligeti, Y. Nir, and H. Quinn, Phys.~Rev. D {\bf 69},
011502(R) (2004).

\bibitem{rpr0_Belle}
  J.~Zhang {\it et al.} (Belle Collaboration), Phys.~Rev.~Lett. {\bf 91}, 221801 (2003).


\bibitem{Belle_Kstarrhop}
J. Zhang {\it et al} (Belle Collaboration), Phys.~Rev.~Lett. {\bf 95},
141801 (2005).







\end{thebibliography}
\end{document}